\documentstyle[12pt]{article}
\input epsf.tex
\newcommand{\beq}{\begin{equation}}
\newcommand{\eeq}{\end{equation}}
\newcommand{\beqa}{\begin{eqnarray}}
\newcommand{\eeqa}{\end{eqnarray}}
\newcommand{\beqar}{\begin{eqnarray*}}
\newcommand{\eeqar}{\end{eqnarray*}}
\renewcommand{\a}{\alpha}
\renewcommand{\b}{\beta}

\newcommand{\e}{\epsilon}
\newcommand{\g}{\gamma}
\newcommand{\G}{\Gamma}

\newcommand{\k}{\kappa}
\renewcommand{\l}{\lambda}

\newcommand{\na}{\nabla}

\newcommand{\z}{\zeta}
\newcommand{\cA}{{\cal A}}

\newcommand{\eg}{{\it e.g.,}\ }
\newcommand{\ie}{{\it i.e.,}\ }
\newcommand{\hphi}{\hat{\phi}}
\newcommand{\ha}{\hat{a}}
\newcommand{\hh}{\hat{h}}
\newcommand{\hD}{d}  %{\hat{D}} %total spacetime dimension
\newcommand{\hd}{\bar{p}} %{\hat{d}} % dimension of transverse subspace
\newcommand{\hg}{\hat{\gamma}}

\newcommand{\pol}{\varepsilon}
\newcommand{\norm}[1]{\raise.3ex\hbox{:}#1\raise.3ex\hbox{:}}
\newcommand{\inn}{\!\cdot\!}
\newcommand{\bz}{\bar{z}}
\newcommand{\bw}{\bar{w}}
\newcommand{\bu}{\bar{u}}
\newcommand{\tV}{\widetilde{V}}
\newcommand{\tG}{\widetilde{G}}
\newcommand{\tS}{\widetilde{S}}
\newcommand{\tX}{\tilde{X}}
\newcommand{\tphi}{\tilde{\phi}}
\newcommand{\tpsi}{\tilde{\psi}}
\newcommand{\labell}[1]{\label{#1}} %{\label{#1}\qquad_{#1}}
\newcommand{\labels}[1]{\label{#1}} %{\vskip-2ex$_{#1}$\label{#1}}
\newcommand{\slf}{\G} %{\hbox{{$F$}\llap{$/$}}}
\newcommand{\Tr}{{\rm Tr}}
\newcommand{\ev}{(\e^v)}
\newcommand{\en}{(\e^n)}

\begin{document}
\begin{titlepage}
\rightline{\small hep-th/9603194 \hfill McGill/96-08}
\vskip 5em

\begin{center}
{\bf \huge Superstring Scattering\\[.25em]
             from D-Branes}
\vskip 3em

{\large Mohammad R. Garousi\footnote{garousi@hep.physics.mcgill.ca}}
and {\large Robert C. Myers\footnote{rcm@hep.physics.mcgill.ca}}
\vskip 1em

{\em	Department of Physics, McGill University \\
	Montr\'eal, Qu\'ebec, Canada H3A 2T8}
\vskip 4em

\begin{abstract}
We derive fully covariant expressions for all two-point scattering 
amplitudes of two massless closed strings from a Dirichlet $p$-brane
This construction relies on the observation that there is a simple
relation between these D-brane amplitudes in type II superstring
theory and four-point
scattering amplitudes for type I open superstrings.
{}From the two-point amplitudes, we derive the long range background
fields for the D-branes, and verify that as expected they correspond
to those of extremally charged $p$-brane solutions of the low
energy effective action.
\end{abstract}
\end{center}

\end{titlepage}

\setcounter{footnote}{0}
\section{Introduction}

Recent exciting progress in string theory has revealed  many
new connections between superstring theories which had previously
been regarded as distinct theories\cite{conn}.
In fact it may be that all string theories are
different phases of a single underlying theory in eleven
or twelve dimensions \cite{fmtheory}. 
Within these discussions, extended
objects, other than just strings, play an important role.
Hence these developments have generated a renewed
interest in $p$-branes (\ie $p$-dimensional extended objects)
and their interactions.

In type II superstring theories, there is a remarkably simple description
of $p$-branes carrying Ramond--Ramond (R-R) charges\cite{joe}. 
The string background is taken to be simply flat empty space,
however interactions of closed superstrings with
these $p$-branes are described by
world-sheets with boundaries fixed to a particular surface at the 
position of a $p$-brane.
The latter is accomplished by imposing Dirichlet boundary conditions
on the world-sheet fields \cite{Dbrane,Dbreview}.
Hence these objects are referred to as Dirichlet $p$-branes
(D$p$-branes) or generically as simply D-branes. 
Within the type IIa theory, the D$p$-branes can have $p=0$, 2, 4, 6
or 8, while for the type IIb strings, $p$ ranges over $-1$, 1, 3, 5, 7,
9 \cite{joe}. So far there have been
only limited results in calculating the scattering amplitudes describing
the interactions of closed strings with D-branes\cite{igora,igorb,form},
and the present paper provides an extension of these
previous works.

The paper is organized as follows: In the following section
we describe the calculation of the scattering of two massless states 
in the Neveu-Schwarz--Neveu-Schwarz (NS-NS) closed string sector from
a Dirichlet $p$-brane using conformal field theory techniques. Our
result extends the calculation of Klebanov and Thorlacius \cite{igora}
to fully covariant amplitudes (without any restrictions on the
polarization tensor). We conclude this section by observing that
the above calculation exactly parallels that of a four-point amplitude
of massless NS states for the open superstring. This observation then
provides a general method for the construction of any two-point
D-brane amplitude. In section \ref{rest} we present all other two-point
amplitudes for scattering from a Dirichlet $p$-brane by using previously
calculated open superstring amplitudes. Our results include the
scattering amplitudes with bosonic R-R states, and
also fermionic NS-R and R-NS states.
In section \ref{back}, we examine the
massless closed string poles in these amplitudes. By comparing these
terms to those in analogous field theory calculations, we are able to
extract the long range background fields surrounding a D$p$-branes.
Our calculations verify that these fields do correspond to those
of extremally charged $p$-brane solutions of the low energy theory.
We conclude with a discussion of our results in section \ref{discuss}.
Appendices \ref{kin} and \ref{spinor} contain some useful information
on the conventions used in our calculations.

\section{NS-NS scattering amplitudes} \labels{nsns}

We begin by calculating the amplitudes describing the
scattering of two massless \hbox{NS-NS} states from a Dirichlet $p$-brane
(\ie the scattering of gravitons, dilatons
or Kalb-Ramond (antisymmetric tensor) states).
The amplitudes are calculated as two closed string vertex operator insertions
on a disk with appropriate boundary conditions
\cite{Dbrane,Dbreview}.
For a D$p$-brane,\footnote{A $p$-brane is an object
extended in $p$ spatial directions which then sweeps out a $p+1$
dimensional world-volume in ten-dimensional Minkowski-signature
spacetime. The special case, $p=-1$, refers to a Euclidean instanton.}
standard Neumann boundary conditions are imposed at the disk boundary
on the world-sheet fields associated with the $p+1$ directions parallel
to the brane's world-volume.
The fields associated with the remaining $9-p$ coordinates orthogonal to
the brane satisfy Dirichlet boundary conditions, which fixes the world-sheet
boundary to the $p$-brane. Using a conformal
transformation, the amplitude can be represented as a 
calculation in the upper half of the complex plane with
the real axis as the world-sheet boundary.
The amplitude may then be written as
\beq
A\simeq\int d^2\!z_1\, d^2\!z_2\ \langle\, V_1(z_1,\bz_1)
\ V_2(z_2,\bz_2)\,\rangle
\labell{ampone}
\eeq
where the vertex operators are
\beqa
V_1(z_1,\bz_1)&=&\pol_{1\mu\nu}\,\norm{V_{-1}^\mu(p_1,z_1)}
\ \norm{\tV_{-1}^\nu(p_1,\bz_1)}
\nonumber\\
V_2(z_2,\bz_2)&=&\pol_{2\mu\nu}\,\norm{V_{0}^\mu(p_2,z_2)}
\ \norm{\tV_{0}^\nu(p_2,\bz_2)}\ \ .
\labell{vertone}
\eeqa
The holomorphic components above are given by
\beqa
V_{-1}^\mu(p_1,z_1)&=&e^{-\phi(z_1)}\,\psi^{\mu}(z_1)\,e^{ip_1\cdot X(z_1)}
\nonumber\\
V_0^\mu(p_2,z_2)&=&\left(\partial X^\mu(z_2)+ip_2\inn \psi(z_2)\psi^{\mu}(z_2)
\right)\,e^{ip_2\cdot X(z_2)}\ \ .
\labell{vertright}
\eeqa
The antiholomorphic components take the same form as in eq.~(\ref{vertright})
but with the left-moving fields replaced by their
right-moving counterparts -- \ie 
$X(z)\rightarrow\tX(\bz)$, $\psi(z)\rightarrow\tpsi(\bz)$, and
$\phi(z)\rightarrow\tphi(\bz)$. As usual, the momenta and polarization
tensors satisfy
\[
p_i^2=0\ ,\qquad\qquad p_i^\mu\,\pol_{i\mu\nu}=0=\pol_{i\mu\nu}\,p_i^\nu\ \ .
\]
and the various physical states would be represented with
\beqa
{\rm graviton}:&&\quad \pol_{i\mu\nu}=\pol_{i\nu\mu},\ \ \pol_{i\mu}{}^\mu=0
\nonumber\\
{\rm dilaton}:&&\quad \pol_{i\mu\nu}=\frac{1}{\sqrt{8}}\left(\eta_{\mu\nu}
-p_{i\mu}\ell_{i\nu}-\ell_{i\mu}p_{i\nu}\right)
\ \ {\rm where}\ p_i\inn\ell_i=1
\nonumber\\
{\rm Kalb-Ramond}:&&\quad \pol_{i\mu\nu}=-\pol_{i\nu\mu}\ \ .
\labell{nspol}
\eeqa
In the amplitude (\ref{ampone}), both integrals run over the upper half
of the complex plane. As a result, this expression is actually divergent
because of the $SL(2,R)$ invariance of the integrand. 
We chose to fix this $SL(2,R)$ invariance
by hand (as opposed to introducing diffeomorphism ghosts)
since we found it to be a useful intermediate check of our
calculations. 

Separately,
the left- and right-moving fields have standard propagators on the
upper half plane, \eg
\beqa
\langle X^{\mu}(z)\,X^{\nu}(w)\rangle&=&-\eta^{\mu \nu}\,\log(z-w)
\nonumber\\
\langle\psi^{\mu}(z)\,\psi^{\nu}(w)\rangle&=&-\frac{\eta^{\mu \nu}}{z-w}
\nonumber\\
\langle\phi(z)\,\phi(w)\rangle&=&-\log(z-w)
\labell{standard}
\eeqa
with analogous expressions for the right-movers
(see  \cite{danf,pkllsw}).
As a result of the boundary at the real axis,
there are also nontrivial correlators between
the right- and left-modes as well
\beqa
\langle X^{\mu}(z)\,\tX^{\nu}(\bw)\rangle&=&-D^{\mu \nu}\,\log(z-\bw)
\labell{xcor}\\
\langle\psi^{\mu}(z)\,\tpsi^{\nu}(\bw)\rangle&=&-\frac{D^{\mu\nu}}{z-\bw}
\labell{psicor}\\
\langle\phi(z)\,\tphi(\bw)\rangle&=&-\log(z-\bw)\ \ .
\nonumber
\eeqa
These propagators have the standard form (\ref{standard})
for the fields satisfying
Neumann boundary conditions, while the matrix $D$ reverses the sign
for the fields satisfying Dirichlet conditions, 
\ie, for $X^\mu$ and $\psi^\mu$ for $\mu=p+1,\ldots,9$
(see eq.~(\ref{dmatrix}) in Appendix \ref{kin}). To simplify the
calculations, we make the replacements
\beq
\tX^\mu(\bz)\rightarrow D^\mu{}_\nu\,X^\nu(\bz)
\qquad
\tpsi^\mu(\bz)\rightarrow D^\mu{}_\nu\,\psi^\nu(\bz)
\qquad
\tphi(\bz)\rightarrow \phi(\bz)
\labell{replace}
\eeq
and which allows us to use the standard correlators (\ref{standard})
throughout our calculations, \ie we extend the fields
to the entire complex plane \cite{igorb}.
With these replacements, the vertex operators (\ref{vertone}) become
\beqar
V_1(z_1,\bz_1)&=&\pol_{1\mu\lambda}D^\lambda{}_\nu\,\norm{V_{-1}^\mu(p_1,z_1)}
\ \norm{V_{-1}^\nu(D\inn p_1,\bz_1)}
\\ 
V_2(z_2,\bz_2)&=&\pol_{2\mu\lambda}D^\lambda{}_\nu\,\norm{V_{0}^\mu(p_2,z_2)}
\ \norm{V_{0}^\nu(D\inn p_2,\bz_2)}
\eeqar
using only the expressions in eq.~(\ref{vertright}).

It is then straightforward to evaluate the correlation function appearing
in the amplitude (\ref{ampone}), and to confirm that the result is
$SL(2,R)$ invariant. To fix this invariance, we set $z_1=iy$ and $z_2=i$.
Introducing the appropriate $SL(2,R)$ Jacobian,\footnote{Our conventions
are such that $z=x+iy$ and $d^2\!z=2dxdy$.}
\[
d^2\!z_1\ d^2\!z_2\rightarrow 4(1-y^2)dy
\]
we are left with a single real integral of the form
\beqa
A&=&-i\kappa T_p\, 2^{p_1\cdot D\cdot p_1+p_2\cdot D\cdot p_2+1}\,
\int_0^1dy\ y^{p_2\cdot D\cdot p_2} (1-y)^{2p_1\cdot p_2}
(1+y)^{2p_1\cdot D\cdot p_2}
\nonumber\\
&&\qquad\qquad\qquad\quad\times\ \left[\frac{1}{1-y^2}a_1-\frac{(1-y)}
{4y(1+y)}a_2\right]
\labell{norm}
\eeqa
where $a_1$ and $a_2$ are two kinematic factors depending only
on the spacetime momenta and polarization tensors.
We have also normalized the amplitude at this
point by the introduction of factors of $\kappa$ and $T_p$, the closed
string and D-brane coupling constants,
respectively.\footnote{Here and in the
subsequent amplitudes, we omit the Dirac delta-function which
imposes momentum conservation in the directions to the $p$-brane
world-volume, \ie eq.~(\ref{conserve}). We have introduced
a phase $-i$ though which corresponds to that of the analogous field
theory amplitudes calculated in Minkowski space ---
see sect.~\ref{massless}.}
The easiest approach to evaluating these integrals is to transform
\[
y=\frac{1-x^{1/2}}{1+x^{1/2}}
\]
which essentially maps the integral to a radial integral on the unit
disk. This transformation has two remarkable effects:
First, the momentum-dependent power of two which appears as 
an overall factor in eq.~\ref{norm} is cancelled using various 
momentum identities (see Appendix \ref{kin}). In particular, one
has momentum conservation in the directions
parallel to the $p$-brane,
\beq
(p_1+D\inn p_1+p_2+D\inn p_2)^\mu=0\ \ .
\labell{conserve}
\eeq
The second effect is that the remaining
integrals take the form of Euler beta functions. Hence,
the final result may be written as
\beq
A=-i\frac{\kappa\,T_p}{2}\,\frac{\G(-t/2)\G(2q^2)}{\G(1-t/2+2q^2)}
\left(2q^2\,a_1+\frac{t}{2}\,a_2\right)
\labell{finone}
\eeq
where $t=-(p_1+p_2)^2=-2p_1\inn p_2$ is the momentum transfer to the
$p$-brane, and $q^2=p_1\inn V\inn p_1={1\over2}p_1\inn D\inn p_1$ is
the momentum flowing parallel to the world-volume of the brane
(see Appendix \ref{kin}). The kinematic factors above are:
\beqa
a_1&=&{\rm Tr}(\pol_1\inn D)\,p_1\inn \pol_2 \inn p_1 -p_1\inn\pol_2\inn
D\inn\pol_1\inn p_2 - p_1\inn\pol_2\inn\pol_1^T \inn D\inn p_1
\nonumber\\
&&\ -p_1\inn\pol_2^T \inn \pol_1 \inn D \inn p_1  
- p_1\inn\pol_2\inn\pol_1^T \inn p_2 +
q^2\,{\rm Tr}(\pol_1\inn\pol_2^T)
+\Big\{1\longleftrightarrow 2\Big\}
\labell{fintwo}\\
\nonumber\\
a_2&=&{\rm Tr}(\pol_1\inn D)\,(p_1\inn\pol_2\inn D\inn p_2 + p_2\inn
D\inn\pol_2\inn p_1 +p_2\inn D\inn\pol_2\inn D\inn p_2)
\nonumber\\
&&+p_1\inn D\inn\pol_1\inn D\inn\pol_2\inn D\inn p_2 -p_2\inn
D\inn\pol_2\inn\pol_1^T\inn D\inn p_1
+q^2\,{\rm Tr}(\pol_1\inn D\inn \pol_2\inn D)
\nonumber\\
&&-q^2\,{\rm Tr}(\pol_1\inn\pol_2^T)
-{\rm Tr}(\pol_1\inn D) {\rm Tr}(\pol_2\inn D)\,(q^2-t/4)
+\Big\{1\longleftrightarrow 2 \Big\}\ \ .
\labell{finthree}
\eeqa
Our notation is such that \eg $p_1\inn\pol_2\inn\pol_1^T \inn D\inn p_1
=p_1^\mu\,\pol_{2\mu\nu}\,\pol_1{}^{\lambda\nu}\,D_{\lambda\rho}\,p_1^\rho$.

{}From the gamma function factors appearing in eq.~(\ref{finone}),
we see that the amplitudes contain two infinite
series poles\footnote{We explicitly restore $\alpha^\prime$ here. Otherwise
our conventions set $\alpha^\prime=2$.}
corresponding to closed string states in the $t$-channel
with $\alpha^\prime m^2=4n$, and to open string states in the $q^2$-
or $s$-channel
with $\alpha^\prime m^2=n$, with $n=0,1,2,\ldots$.
As is evident the final amplitude is symmetric under the interchange
of the two string states, \ie $1\longleftrightarrow 2$, despite the
asymmetric appearance of the initial integrand in eq.~(\ref{ampone}).
Another check is that the amplitude satisfies the Ward identities
associated with the gauge invariances of these states, \ie the
amplitude vanishes upon substituting $\pol_{i\mu\nu}\rightarrow
p_{i\mu}\, q_{i\nu}$ or $q_{i\mu}\, p_{i\nu}$, where $q_i\cdot p_i=0$.
In the special case that the polarization tensors have non-vanishing
components only in directions perpendicular to the world-volume
of the $p$-brane (\ie following Appendix \ref{kin}, 
$V_\mu{}^\nu\,\pol_{i\nu\rho}=0=\pol_{i\mu\nu}\, V^\nu{}_\rho$)
the results simplify greatly. In this
case, we find for two gravitons
\[
A=-i\frac{\kappa\,T_p}{2}\,\frac{\G(-t/2)\G(1+2q^2)}{\G(1-t/2+2q^2)}
\,2q^2\,{\rm Tr}(\pol_2\inn\pol_1)
\]
while for two Kalb-Ramond particles,
\[
A=-i\frac{\kappa\,T_p}{2}\,\frac{\G(-t/2)\G(1+2q^2)}{\G(1-t/2+2q^2)}
\left(4\, p_1\inn\pol_2\inn\pol_1\inn p_2+(t-2q^2)\,
{\rm Tr}(\pol_2\inn\pol_1)\right)
\]
which agrees with the results of Ref.~\cite{igora,igorb}.

The purpose behind our rather lengthy description of the calculations
for the NS-NS scattering amplitudes is to compare these calculations
to that of an apparently unrelated scattering amplitude, namely, 
the amplitude for four
massless NS vectors in open superstring theory. The latter would
be calculated as four vertex operator insertions on the boundary of
a disk on which Neumann boundary conditions are imposed. Again the
disk can be mapped to the upper half of the complex plane, in which case
the amplitude becomes
\beqa
A&\simeq&\int dx_1\,dx_2\,dx_3\,dx_4 \langle\, \norm{\z_1\cdot V_{-1}(2k_1,x_1)}
\ \norm{\z_2\cdot V_{0}(2k_2,x_2)}
\nonumber\\
&&\qquad\qquad\times\ \norm{\z_3\cdot V_{0}(2k_3,x_3)}
\ \norm{\z_4\cdot V_{-1}(2k_4,x_4)}\,\rangle
\labell{ampopen}
\eeqa
where the vertex operators are written in terms of the same components 
given in eq.~(\ref{vertright}) which were used to construct the previous
closed string amplitudes.\footnote{Each open string vertex operator is
labelled with momentum
$2k_i$ since we maintain our convention that $\alpha^\prime=2$ even though
this is an open string scattering amplitude.}
One evaluates the correlation functions using the same standard propagators
appearing in eq.~(\ref{standard}). In this amplitude (\ref{ampopen}), 
$x_i$ lie on the real axis and the range of integration
is $-\infty\le x_1\le x_2\le x_3\le x_4 \le\infty$.
Therefore this expression diverges because of the $SL(2,R)$
invariance of the integrand. The standard approach to fixing this
invariance is to set $x_1=0,x_2=x,x_3=1,x_4=\infty$. Here,
we make an alternate choice instead setting $x_1=-1,$ $x_2=-x_3$,
$x_3=x$, and $x_4=1$. Introducing the appropriate $SL(2,R)$ Jacobian,
\[
dx_1\,dx_2\,dx_3\,dx_4 \rightarrow (1-x^2)dx
\]
we are left with a single real integral
\[
A\simeq\int_0^1dx\ x^{4k_2\cdot k_3} (1-x)^{8k_1\cdot k_2}
(1+x)^{8k_1\cdot k_3}
\left[\frac{1}{1-x^2}a'_1-\frac{1-x}{4x(1+x)}a'_2\right]
\]
which has essentially the same form as eq.~(\ref{norm}) for the
closed string amplitudes.
The final result here is well-known \cite{jhsreport} and may be written
as\footnote{Here and in the following section, we present the open superstring
amplitudes as calculated with the conformal field theory conventions
described above --- see also Appendix \ref{spinor} for spin operators.}
\beq
A(\z_1,k_1;\z_2,k_2;\z_3,k_3;\z_4,k_4)=-\frac{1}{2}g^2
\frac{\G(4k_1\inn k_2)\G(4k_1\inn k_4)}{\G(1+4k_1\inn k_2
+4k_1\inn k_4)}K(\z_1,k_1;\z_2,k_2;\z_3,k_3;\z_4,k_4)\ \ .
\labell{openresult}
\eeq
where $g$ is the open string coupling constant. The (rather lengthy)
kinematic factor may be written as
\beqa
K&=&-16k_2\inn k_3\,k_2\inn k_4\ \z_1\inn\z_2\,\z_3\inn\z_4
\nonumber\\
&&\ \ -4k_1\inn k_2\,(\z_1\inn k_4\,\z_3\inn k_2\,\z_2\inn\z_4
+\z_2\inn k_3\,\z_4\inn k_1 \,\z_1\inn\z_3+\z_1\inn k_3\,\z_4\inn
k_2\,\z_2\inn\z_3 +\z_2\inn k_4\,\z_3\inn k_1\,\z_1\inn\z_4)
\nonumber\\
&&\ \ \ \ +\Big\{1,2,3,4\rightarrow 1,3,2,4\Big\}+
\Big\{1,2,3,4\rightarrow 1,4,3,2\Big\} \ \ .
\nonumber
\eeqa
One can see that this expression displays an infinite set of open
string poles in both the $s$- and $t$-channels with $\alpha^\prime 
m^2=n$ where $n=0,1,2,\ldots$.

Now what becomes clear from this discussion is that both calculations
involve precisely the same correlation functions and 
exactly the same integrals after fixing the $SL(2,R)$ invariance. Thus up to
an overall normalization of the coupling constants, both amplitudes are
identical. One need only make the following substitution to convert the
four vector amplitude in the open superstring theory
into a D-brane amplitude for two NS-NS closed superstring states:
\beqa
2k_1^\mu\rightarrow p_1^\mu&&2k_4^\mu\rightarrow (D\inn p_1)^\mu
\nonumber\\
2k_2^\mu\rightarrow p_2^\mu&&2k_3^\mu\rightarrow (D\inn p_2)^\mu
\nonumber\\
\z_{1\mu}\,\otimes\,\z_{4\nu}&\rightarrow \pol_{1\mu\lambda}D^\lambda{}_\nu&
\nonumber\\
\z_{2\mu}\,\otimes\,\z_{3\nu}&\rightarrow \pol_{2\mu\lambda}D^\lambda{}_\nu
& ,
\labell{transone}
\eeqa
as well as replacing $g^2\rightarrow i\kappa\,T_p$.
We have directly confirmed that these replacements in eq.~(\ref{openresult})
precisely reproduce the results in
eqs.~(\ref{finone}-\ref{finthree}). It is interesting to
note that with these substitutions momentum conservation in the
open string amplitude becomes precisely momentum conservation
in the directions parallel to the D-brane as in eq.~(\ref{conserve}). 
These observations above allow us
to easily calculate any Dirichlet two-point amplitudes by simply
using the well-known results for four-point open string scattering
amplitudes \cite{jhsreport}, as described in the following section.

\section{Dirichlet two-point amplitudes} \labels{rest}

The relation between the four-vector scattering amplitude
for the open superstring and the two-point amplitude for scattering
of two NS-NS closed superstring states from a Dirichlet brane may
seem somewhat surprising. In fact, it might be regarded as an extension
of the results in Ref.~\cite{klt}. There it was shown that closed
string amplitudes could be expressed as products of open string amplitudes
corresponding to the correlation functions appearing in the independent
right- and left-moving sectors along with certain ``sewing'' factors.
The present case is similar except that with the D-brane
boundary conditions,
the right- and left-movers are naturally ``sewn'' together in a single
open string amplitude. Given this idea, it is straightforward to
write down all of the remaining Dirichlet brane two-point amplitudes
using the well-known four-particle open superstring amplitudes
involving massless spinors, as well as vectors.
These amplitudes may all be expressed in the form \cite{jhsreport}
\beq
A(1,\,2,\,3,\,4)=-\frac{1}{2}g^2
\frac{\G(4k_1\inn k_2)\G(4k_1\inn k_4)}{\G(1+4k_1\inn k_2+4k_1\inn k_4)}
K(1,\,2,\,3,\,4)\ \ .
\labell{opentot}
\eeq
The various kinematic factors are then given by
\beqa
K(u_1,\,u_2,\,u_3,\,u_4)&=&
-2\,k_1\inn k_2\,\bu_2 \g^\mu u_3\,\bu_1\g_\mu u_4
+2\,k_1\inn k_4\,\bu_1\g^\mu u_2\,\bu_4\g_\mu u_3
\labell{openone}\\
K(u_1,\,\z_2,\,\z_3,\,u_4)&=&
2i\sqrt{2}\,k_1\inn k_4\,\bu_1\g\inn \z_2\g\inn(k_3+k_4)\g\inn\z_3 u_4
\labell{opentwo}\\
&&-4i\sqrt{2}\,k_1\inn k_2\,\left(\bu_1\g\inn\z_3u_4\,k_3\inn\z_2
-\bu_1\g\inn\z_2u_4\,k_2\inn\z_3-\bu_1\g\inn k_3u_4\,\z_2\inn\z_3\right)
\nonumber\\
K(u_1,\,\z_2,\,u_3,\,\z_4)&=&
-2i\sqrt{2}\,k_1\inn k_4\,\bu_1\g\inn \z_2\g\inn(k_3+k_4)\g\inn\z_4 u_3
\labell{openthree}\\
&&\qquad\quad\qquad-2i\sqrt{2}\,k_1\inn k_2\,\bu_1\g\inn
\z_4\g\inn(k_2+k_3)\g\inn\z_2 u_3
\ \ .
\nonumber
\eeqa
In translating these results to Dirichlet two-point amplitudes, schematically
ones associates (some cyclic permutation of) $(1,2,3,4)$ in the open string
amplitude with $(1_L,2_L,2_R,$ $1_R)$ in the closed string amplitude,
where here the subscripts $L$ and $R$ denote the left- and right-moving
components of the closed string states. Having chosen a particular
ordering, the translation of the NS sector contributions between
the open and closed string amplitudes remains the same as
in eq.~(\ref{transone}). The 
interesting feature is the appearance of a factor of $D_{\mu\nu}$ in the
momenta and polarization tensors of the right-moving contributions.
An analogous spinor matrix $M_{AB}$
also appears in the right-moving Ramond sector contributions
--- see Appendix \ref{spinor}.
The D-brane scattering amplitudes then take a universal form
\beq
A(1,\,2)=-i\,\frac{\kappa\,T_p}{2}\,\frac{\G(-t/2)\G(2q^2)}{\G(1-t/2+2q^2)}
K(1,\,2)
\labell{ampuniverse}
\eeq
where as before $t$ is the momentum transfer to the
$p$-brane, and $q^2$ is
the momentum flowing parallel to the world-volume of the brane
--- see Appendix \ref{kin}. For later discussions, it is 
useful to divide the kinematic factor as
\beq
K(1,\,2)=2q^2\,a_1(1,\,2)+\frac{t}{2}\,a_2(1,\,2)
\labell{subdivide}
\eeq
as was done in eq.~(\ref{finone}). Then $a_1(1,\,2)$ will be essentially
the residue of the massless $t$-channel pole, which will become
important for the analysis in sect.~\ref{massless}.
Now, it simply remains to translate
the kinematic factors (\ref{openone})-(\ref{openthree}) in the
appropriate way.

\subsection{R-R boson amplitude}

The simplest case is using eqs.~(\ref{opentot}) and (\ref{openone}) to
calculate the amplitude describing two R-R states scattering from a
Dirichlet brane. The latter amplitude would be written as
\[
A\simeq \int d^2\!z_1\, d^2\!z_2\ \langle\, V_1(z_1,\bz_1)
\ V_2(z_2,\bz_2)\,\rangle
\]
where the vertex operators are
\beq
V_i(z_i,\bz_i)=(P_-\,\slf_{i(n)})^{AB}\,\norm{V_{-1/2\,A}(p_i,z_i)}
\ \norm{\tV_{-1/2\,B}(p_i,\bz_i)}\ \ .
\labell{vertspin}
\eeq
The holomorphic components above are given by
\beq
V_{-1/2\,A}(p_i,z_i)=e^{-\phi(z_i)/2}\,S_A(z_i)\,e^{ip_i\cdot X(z_i)}
\labell{vertspinor}
\eeq
and the antiholomorphic components have the same form, but with the
left-moving fields replaced by their right-moving counterparts. As before
we use eq.~(\ref{replace}) to replace $\tX^\mu$ and $\tphi$ in $\tV_{-1/2\,B}$.
Similarly, the right-moving spin field is replaced using \cite{igorb}
(see also \cite{gut})
\beq
\tS_A(\bz)\rightarrow M_A{}^B\, S_B(\bz)
\labell{replaced}
\eeq
where $M_{AB}$ is defined in Appendix \ref{spinor}.
With this replacement, only standard correlators of the
spin fields \cite{fms,danf,pkllsw} appear in the subsequent calculations.
We have explicitly included the chiral projection operator
$P_-=(1-\g_{11})/2$ in vertex operator (\ref{vertspin}),
so that our calculations are always made with the full 32$\times$32 Dirac
matrices of ten dimensions. We have also defined
\beq
\slf_{i(n)}=\frac{a_n}{n!}F^i_{\mu_1\cdots\mu_n}\,\g^{\mu_1}\cdots
\g^{\mu_n}\ \ .
\labell{self}
\eeq
With our choice of conventions (see Appendix \ref{spinor}), we
must introduce the factor $a_n=i$ for 
the $n=2$ and 4 fields in the type IIa theory, while
$a_n=1$ for $n=1$, 3 and 5 in the type IIb theory. In eq.~(\ref{self}),
$F^i_{\mu_1\cdots\mu_n}$ is the linearized $n$-form field strength
with
\beqa
F^i_{\mu_1\cdots\mu_n}&=&i\, n\, p_{i[\mu_1}\pol_{i\mu_2\cdots\mu_n]}
\labell{strong}\\
&=&i\,p_{i\mu_1}\pol_{i\mu_2\cdots\mu_n} \pm\ {\rm cyclic\ permutations}
\nonumber
\eeqa
where $p_i^2=0$ and $p_i^\mu\,\pol_{i\mu\mu_3\cdots\mu_n}=0$.
Hence the appropriate substitutions for the open string
amplitude (\ref{openone}) to derive the Dirichlet amplitude are
\beqa
2k_1^\mu\rightarrow p_1^\mu&&2k_4^\mu\rightarrow (D\inn p_1)^\mu
\nonumber\\
2k_2^\mu\rightarrow p_2^\mu&&2k_3^\mu\rightarrow (D\inn p_2)^\mu
\nonumber\\
u_{1A}\,\otimes\,u_{4B}&\rightarrow (P_-\,\slf_{1(n)}M)_{AB}&
\nonumber\\
u_{2A}\,\otimes\,u_{3B}&\rightarrow (P_-\,\slf_{2(m)}M)_{AB}
& .
\labell{transtwo}
\eeqa
The resulting kinematic factor is:
\beqa
K_{R-R,R-R}&=&
-q^2\,\Tr(P_-\slf_{1(n)}M\g_\mu C^{-1} M^T\slf^T_{2(m)}C\g^\mu)
\nonumber\\
&&\qquad+\frac{t}{4}\Tr(P_-\slf_{1(n)}M\g_\mu)\,\Tr(P_-\slf_{2(m)}M\g^\mu)
\labell{kinrrrr}
\eeqa
This scattering amplitude was previously calculated in Ref.~\cite{igorb}.
In their results, $a_1$ (\ie the first term above) does not
quite appear with the same form as here. However, 
in sect.~\ref{massless}
with an explicit evaluation of the trace in $a_1$,
we will show that our results are identical to those of \cite{igorb}.

\subsection{NS-NS and R-R amplitude}

The next case is calculating the amplitude describing one R-R and 
one NS-NS state scattering from a Dirichlet brane
using eqs.~(\ref{opentot}) and (\ref{opentwo}).
The appropriate substitutions to derive the Dirichlet amplitude are already
derived for the previous amplitudes in eqs.~(\ref{transone})
and (\ref{transtwo})
\beqar
2k_1^\mu\rightarrow p_1^\mu&&2k_4^\mu\rightarrow (D\inn p_1)^\mu
\\ 
2k_2^\mu\rightarrow p_2^\mu&&2k_3^\mu\rightarrow (D\inn p_2)^\mu
\nonumber\\
u_{1A}\,\otimes\,u_{4B}&\rightarrow (P_-\,\slf_{1(n)}M)_{AB}&
\\ 
\z_{2\mu}\,\otimes\,\z_{3\nu}\hphantom{B}
&\rightarrow \pol_{2\mu\lambda}D^\lambda{}_\nu\ \ 
\hphantom{M)_{AB}}\ \ .& 
\eeqar
The resulting kinematic factor is then:
\beqa
K_{R-R,NS-NS}&=&
i\frac{q^2}{\sqrt{2}}\Tr[P_-\slf_{1(n)}M\g^\nu\g\inn(p_1+p_2)\g^\mu]\,(
\pol_2\cdot D)_{\mu\nu}
\nonumber\\
&&\ \ -i\frac{t}{2\sqrt{2}}\left[\Tr(P_-\slf_{1(n)}M\g\inn
D\inn\pol^T_2\inn D\inn p_2)
-\Tr(P_-\slf_{1(n)}M\g\inn \pol_2\inn D\inn p_2)\right.
\nonumber\\
&&\left.\ \ 
\qquad\qquad-\Tr(P_-\slf_{1(n)}M\g\inn D\inn p_2)\,\Tr(\pol_2\inn D)\right]
\labell{kinrrnn}
\eeqa
Just as for eq.~(\ref{finone}),
this result inherits the gauge invariance of the NS states in the
open string amplitude as the closed string gauge invariance for
the NS-NS state. This amplitude will be of particular interest
in the following section for determining the background R-R fields
in sect.~\ref{back}.

\subsection{R-NS fermion amplitude}

We can also calculate the fermionic scattering amplitudes
as well. Making the alternative match in eq.~(\ref{opentwo}) which
identifies $(4,1,2,3)$ with $(1_L,2_L,2_R,1_R)$ results in a scattering
amplitude describing two states in the R-NS sector of the closed string.
The latter amplitude would be written as
\[ 
A\simeq \int d^2\!z_1\, d^2\!z_2\ \langle\, V_1(z_1,\bz_1)
\ V_2(z_2,\bz_2)\,\rangle
\] 
where the vertex operators are
\beqar
V_1(z_1,\bz_1)&=&P_-^{AB}\psi_{1\mu B} D^\mu{}_\nu\,
\norm{V_{-1/2\,A}(p_1,z_1)}\ \norm{V_{-1}^\nu(D\inn p_1,\bz_1)}
\\ 
V_2(z_2,\bz_2)&=&P_-^{AB}\psi_{2\mu B}D^\mu{}_\nu\,
\norm{V_{-1/2\,A}(p_2,z_2)}\ \norm{V_0^\nu(D\inn p_1,\bz_1)}
\eeqar
where $V_0$ and $V_{-1}$ are given eq.~(\ref{vertright}), while
$V_{-1/2}$ appears in eq.~(\ref{vertspinor}). Here we have already
replaced the right-moving modes by the appropriate expressions as
in eqs.~(\ref{replace}) and (\ref{replaced}).
Again we have explicitly included the chiral projection operator
$P_-=(1-\g_{11})/2$, even though the polarization tensors implicitly
satisfy $P_-\psi_{i\mu}=\psi_{i\mu}$. Further
the momenta and polarizations satisfy
\[
{p_i}^2=0\ \qquad\qquad(\g\inn p_i)_A{}^B\,\psi_{i\mu B}=0\
\qquad\qquad p^{\mu}_{i}\psi_{i\mu B}=0\ \ .
\]
The physical spinor states are represented with
\beqar
{\rm gravitino}:&\quad \psi_{i\mu A}\qquad&\ {\rm where}\ 
(\g^\mu)_A{}^B\,\psi_{i\mu B}=0
\\
{\rm dilatino}:&\quad \psi_{i\mu A}=(\g_\mu)_A{}^B\chi_{iB}
&\ {\rm where}\ (\g\inn p_i)_A{}^B\, \chi_{iB}=0\ \ .
\eeqar
The appropriate
substitutions to derive the Dirichlet amplitude are then
\beqar
2k_4^\mu\rightarrow p_1^\mu&&2k_3^\mu\rightarrow (D\inn p_1)^\mu
\\ 
2k_1^\mu\rightarrow p_2^\mu&&2k_2^\mu\rightarrow (D\inn p_2)^\mu
\\ 
u_{4A}\,\otimes\,\z_{3\mu}&\rightarrow (P_-\,\psi_{1}\inn D)_{\mu A}&
\nonumber\\
u_{1A}\,\otimes\,\z_{2\mu}&\rightarrow (P_-\,\psi_{2}\inn D)_{\mu A}
& .
\eeqar
The resulting kinematic factor is then:
\beqa
K_{R-NS,R-NS}&=&
i\sqrt{2}\,q^2(\psi_{2}\cdot p_{1}\g\inn D\inn P_-\psi_{1}-\psi_{2}\cdot
D\inn \g\, p_{2}\inn P_-\psi_{1}
\nonumber\\
&&\qquad\qquad-{\psi_{2}}^{\mu}\,\g\inn D \inn p_{1}\,P_-\psi_{1\mu})
\nonumber\\
&&\  \ +i\frac{t}{4\sqrt{2}}\left(\psi_{2}\cdot D\inn\g\,\g\inn(p_{1}+D\inn
p_{1})\,\g\inn D\inn P_-\psi_{1}\right)
\labell{kinrnrn}
\eeqa
These amplitudes vanish with the substitution
$\psi_{i\mu A}\rightarrow p_{i\mu}\chi_A$ where
$(\g\inn p_i)_A{}^B\, \chi_B=0$. This Ward identity for the gauged
supersymmetry in the closed superstring theory is again naturally
inherited from the vector gauge invariance of the corresponding
open superstring amplitude.

\subsection{NS-R fermion amplitude}

Identifying the cyclic permutation $(2,3,4,1)$ in eq.~(\ref{opentwo}) 
with $(1_L,2_L,2_R,1_R)$ in the Dirichlet amplitude results in a scattering
amplitude for two spinors in the NS-R sector. In this case, the previous
discussion motivates the substitutions
\beqar
2k_2^\mu\rightarrow p_1^\mu&&2k_1^\mu\rightarrow (D\inn p_1)^\mu
\\ 
2k_3^\mu\rightarrow p_2^\mu&&2k_4^\mu\rightarrow (D\inn p_2)^\mu
\\ 
\z_{3\mu}\,\otimes\,u_{4A}&\rightarrow (MP_\pm\,\psi_{2})_{\mu A}&
\\ 
\z_{2\mu}\,\otimes\,u_{1A}&\rightarrow (MP_\pm\,\psi_{1})_{\mu A}
&
\eeqar
where $P_{+(-)}$ is chosen in the type IIa(b) superstring
theory.\footnote{Our convention is that the Ramond ground-state in the
left-moving sector has negative chirality in both theories. For the
right-movers, the opposite (same) chirality is chosen in the type
IIa(b) theory.} The resulting kinematic factor is then:
\beqa
K_{NS-R,NS-R}&=&
i\sqrt{2}\,q^2(\psi_1\inn p_2\,M^{-1}\,\g\inn
M\,P_\pm\psi_1-\psi_2\,M^{-1}\inn\g\, M\,P_{\pm}\,p_1\inn\psi_2
\nonumber\\
&&\qquad\qquad-{\psi_1}^{\mu}\,M^{-1}\,\g\inn p_2\,M\,P_\pm\psi_{2\mu})
\nonumber\\
&&\ \ +i\frac{t}{4\sqrt{2}}(\psi_1\,M^{-1}\inn\g\,\g\inn (p_2+D\inn p_2)
\,\g\inn M\,P_\pm\psi_2)
\labell{kinnrnr}
\eeqa
where $M^{-1}=C^{-1}M^T C$. This kinematic factor again satisfies the
appropriate Ward identities
as in eq.~(\ref{kinrnrn}).
Using identities such as $(M\,\g^{\mu})=D_{\mu\nu}(\g_{\nu}\,M)$
(see Appendix \ref{spinor}),
one can show the two kinematic factors in eqs.~(\ref{kinrnrn})
and (\ref{kinnrnr}) are in fact identical (up to a sign and the chiral
projection in the type IIa theory).

\subsection{R-NS and NS-R amplitude}

Finally eq.~(\ref{openthree}) yields the kinematic factor for the
scattering two fermions with one each  in the NS-R and R-NS sectors.
With
\beqar
2k_1^\mu\rightarrow p_1^\mu&&2k_4^\mu\rightarrow (D\inn p_1)^\mu
\\ 
2k_2^\mu\rightarrow p_2^\mu&&2k_3^\mu\rightarrow (D\inn p_2)^\mu
\\ 
u_{1A}\,\otimes\,\z_{4\mu}&\rightarrow (P_-\,\psi_{1}\cdot D)_{\mu A}&
\\ 
\z_{2\mu}\,\otimes\,u_{3A}&\rightarrow (MP_\pm\,\psi_{2})_{\mu A}
\hphantom{D}
& .
\eeqar
one finds that
\beqar
K_{R-NS,NS-R}&=&
-i\frac{q^2}{\sqrt{2}}(\psi_1\,P_+\cdot
D_{\nu}\,\g^{\mu}\,\g\inn(p_1+p_2)\,\g^{\nu}\,M\,{\psi_2}_{\mu})
\\
&&-i\frac{t}{4\sqrt{2}}(\psi_1\,P_+\cdot D\inn\g\,\g\inn(p_2+D\inn p_2)
\,\g\inn M\,\psi_2)\ \ .
\eeqar
Note that the chiral projections of the spinors are consistent 
\ie allow for a nonvanishing amplitude, because of the form of
$M$ given in eq.~(\ref{finalm}) and the fact that $p$ is even and
odd in the type IIa and IIb theories, respectively.

\section{Background fields} \labels{back}

In the context of various field theories,
many different $p$-brane solutions have been constructed
describing extended objects of different dimensions.
(see \cite{newsol} as well as \cite{report} and references therein).
Typically these solutions involve a $(n-1)$-form potential coupled
to gravity and a scalar field, \ie the dilaton. 
For a potential of form degree $(n-1)$ in $\hD$ spacetime dimensions,
one naturally finds two dual classes of 
$p$-branes. The first with $p=n-2$ carries an ``electric'' charge
of the $n$-form field strength. The dual
object with $p=\hD-n-2$ carries an analogous ``magnetic'' charge.

In the following section, we review some of these $p$-brane
solutions. Then in sect.~\ref{massless}, we examine our D-brane
scattering amplitudes for massless $t$-channel poles. These
poles correspond to the interactions produced by the long
range background fields around the Dirichlet $p$-branes. Finally
in sect.~\ref{compare}, we make a detailed comparison of the
D$p$-brane fields with those in the field theory solutions.

\subsection{Extremal $p$-branes} \labels{lowsol}

We begin with an action in $\hD$ dimensions
\beq
\hat{I}= {1\over16\pi \hat{G}_N}
\int d^{\hD}\!x\,\sqrt{-\hat{g}}\left [\hat{R}(\hat{g})-
{\hg\over 2}(\nabla\hphi)^2
-{1\over 2\,n!}e^{-\ha\hg \hphi}\hat{F}_{(n)}^2 \right]
\labell{actors}
\eeq
describing a potential $\hat{A}$ coupled to gravity and a dilaton.
The field strength $\hat{F}_{(n)}$ is an $n$-form given by
$\hat{F}_{(n)}=d\,\hat{A}$,
and $\hg=2/(\hD-2)$ is a convenient normalization factor.
The focus of our discussion will be solutions describing static, isotropic
and extremal $p$-branes. Here, ``isotropic'' means 
that the solutions are Lorentz invariant in the
directions parallel to the world-volume of the $p$-brane,
while ``extremal'' requires a zero-force condition is satisfied, 
\ie there is a precise cancellation of the static
forces between the $p$-branes
generated by the dilaton, form-field and graviton.
In the case that the action (\ref{actors})
can be embedded as part of a supersymmetric theory,
extremality becomes the condition that the 
$p$-brane solutions themselves are supersymmetric \cite{report}. 
It is because one expects both properties to hold for the configurations
corresponding to D-branes that we restrict the following discussion
to this class of solutions.

We begin with an electrically charged $p$-brane solution
with $p=n-2$. The metric takes the following form:
\beq
d\hat{s}^2=H^{2\a}(\vec{x})\,(-dt^2+d\vec{y}^2)
+H^{2\b}(\vec{x})\,d\vec{x}^2\ \ .
\labell{pmetric}
\eeq
Here the time, $t$, and the $p=n-2$ spatial coordinates, $y^a$, run
parallel to
the surface of the brane, while the orthogonal subspace is covered
by the $\hd=\hD-n+1$ coordinates, $x^i$.
The $(n-1)$-form potential and the dilaton may be
written as
\beq
\hat{A}=\pm \sqrt{2\sigma}\,H(\vec{x})^{-1}\ \e^v
\qquad{\rm and}\qquad e^{-\hphi}=H(\vec{x})^\tau
\labell{pfields}
\eeq
where $\e^v=dt\,dy^1\cdots dy^p$ is the volume form in the
subspace parallel to the $p$-brane world-volume -- see Appendix
\ref{kin}. For later discussion, it is also convenient to
present the field strength $F_{(n)}$. Given the potential in
eq.~(\ref{pfields}), the latter is given by
\[ 
\hat{F}_{(n)}=\mp\sqrt{2\sigma}\,H^{-2}\,\partial_j H\,dx^j\wedge\e^v
\ \ .
\] 

With an appropriate choice of exponents, namely,
\beqa
\alpha&=-\frac{\hd-2}{(p+1)(\hd-2)+\ha^2} 
\qquad\beta&=\frac{p+1}{(p+1)(\hd-2)+\ha^2}
\nonumber\\
\sigma&=\frac{\hD-2}{(p+1)(\hd-2)+\ha^2}\qquad \tau&=\ha\,\sigma
\labell{exponents}
\eeqa
one finds that $H$ should satisfy the flat-space
Laplace's equation in the transverse space, 
\ie $\delta^{ij}\partial_i\partial_j H=0$. Extremal $p$-branes
are constructed by introducing $\delta$-function sources in the
latter equation.
For a single $p$-brane solution as will be of interest in the
following, we choose $H=1+\mu G(|\vec{x}|/\ell)$ with
\beq
G=\left\lbrace\matrix{&\frac{1}{\hd-2}\,\left(\ell/|\vec{x}|\right)^{\hd-2}
&\qquad{\rm for}\ \hd\ge3\cr
&-\log\left(|\vec{x}|/\ell\right)&\qquad{\rm for}\ \hd=2\cr
&-|\vec{x}|/\ell\hphantom{\log()}&\qquad{\rm for}\ \hd=1\cr}\right.
\labell{green}
\eeq
Here $\ell$ is an arbitrary length and $\mu$ is some dimensionless constant.
We will consider the latter
constant to be small, {\it i.e.,} $\mu<<1$, so that we may treat the
nontrivial part of the solutions as a perturbation of flat empty
space. Certainly this is valid in the region $|\vec{x}|>>\ell$ for
$\hd\ge3$, but only a formal expansion for $\hd=2$ and 1.
Note that these solutions may be extended to an instanton with $p=-1$
by using a euclidean
metric without $t$ or $y^a$, and having $\hat{A}$ be a 0-form or scalar
\cite{instanton,igorb}.

One can also construct a magnetically charged $p$-brane solution
with $p=\hD-n-2$. With this choice of $p$ and hence $\hd=n+1$,
the metric remains as given in eq.~(\ref{pmetric}). The dilaton takes
the same form as in eq.~(\ref{pfields}). but with the opposite sign,
\ie $e^{+\hphi}=H(\vec{x})^\tau$. Finally the form field is magnetic
with nonvanishing components in the transverse subspace. The latter
is conveniently expressed in terms of the field strength as
\[ 
\hat{F}_{(n)}=\mp\sqrt{2\sigma}\,\partial_j H\,i_{\hat{x}^j}\e^n
\] 
where $i_{\hat{x}^j}$ denotes the interior product with a unit vector
pointing in the $x^j$ direction. Also $\e^n=dx^1\cdots dx^{\hd}$
is the volume form in the subspace orthogonal to the $p$-brane.
 With these choices, the numerical
factors (\ref{exponents}) are left unchanged when written in terms of
$p$, $\hd$ and $\hD$.

The important feature of these solutions for the purposes
of comparison with the string scattering amplitudes
will be the behavior of fields in the asymptotic region, 
\ie $|\vec{x}|\rightarrow\infty$. Expanding with $\mu<<1$, the
asymptotic metric is essentially flat with
\beqa
\hat{h}_{\mu\nu}&\equiv&\hat{g}_{\mu\nu}-\eta_{\mu\nu}
\labell{asymetric}\\
&\simeq&2\mu\,G(|\vec{x}|/\ell)\,{\rm diag}(-\alpha , \alpha ,\ldots,
         \alpha , \beta ,\ldots,\beta)\ \ .
\nonumber
\eeqa
For the ``electric''  $p$-brane with $p=n-2$, one has
\beqa
\hphi&\simeq&-\ha\,\sigma\,\mu\, G(|\vec{x}|/\ell)
\nonumber\\
\hat{F}_{(n)}&\simeq&\mp\sqrt{2\sigma}\,\mu\,\partial_j G(|\vec{x}|/\ell)
\,dx^j\wedge\e^v\ \ .
\labell{elecfields}
\eeqa
Similarly for the ``magnetic'' $p$-brane with $p=\hD-n-2$,
one finds
\beqa
\hphi&\simeq&+\ha\,\sigma\,\mu\, G(|\vec{x}|/\ell)
\nonumber\\
\hat{F}_{(n)}&\simeq&\mp\sqrt{2\sigma}\,\mu \,\partial_j G(|\vec{x}|/\ell)
\,i_{\hat{x}^j}\e^n\ \ .
\labell{magfields}
\eeqa
Given these results, it is straightforward to derive some of the
physical quantities which characterize these solutions.
The ADM mass per unit $p$-volume is
defined as \cite{massy}:
\[
M_p
={1\over16\pi \hat{G}_N}\oint \sum_{i=1}^{\hd}\ n^i\left[
\sum_{j=1}^{\hd}(\partial_j\hh_{ij}-\partial_i\hh_{jj})-
\sum_{a=1}^p\partial_i\hh_{aa}\right] r^{\hd-1} d\Omega
\]
where $n^i$ is a radial unit vector in the transverse subspace,
and $\hat{G}_N$ is the Newton's constant appearing in the action
(\ref{actors}). Given eqs.~(\ref{exponents}), (\ref{green})
and (\ref{asymetric}), one finds
\beq
M_p=\frac{\sigma}{8\pi \hat{G}_N}\cA_{\hd-1}\,\mu\,\ell^{\hd-2}
\labell{mess}
\eeq
where $\cA_{\hd-1}=2\pi^{\hd}/\Gamma(\hd/2)$
is the area of a unit $(\hd-1)$-sphere.
The electric form charge is given by \cite{report}
\beqa
Q_E&=&\frac{1}{\sqrt{16\pi\hat{G}_N}}\oint *\hat{F}_{(n)}
\nonumber\\
&=&\pm(-)^{p\hd}
\sqrt{\frac{\sigma}{8\pi\hat{G}_N}}\cA_{\hd-1}\,\mu\,\ell^{\hd-2}\ \ .
\labell{elecharge}
\eeqa
Similarly the magnetic form charge is give by
\beqa
Q_M&=&\frac{1}{\sqrt{16\pi\hat{G}_N}}\oint \hat{F}_{(n)}
\nonumber\\
&=&\pm\sqrt{\frac{\sigma}{8\pi\hat{G}_N}}\cA_{\hd-1}\,\mu\,\ell^{\hd-2}\ \ .
\labell{magcharge}
\eeqa

\subsection{Massless $t$-channel poles} \labels{massless}

Recall that in the scattering amplitudes, the momentum transfer
to the D-brane is $t=-(p_1+p_2)^2$. Given the general form
of the string amplitudes in eqs.~(\ref{ampuniverse}-\ref{subdivide}),
one can expand these amplitudes as an infinite sum of terms reflecting
the infinite tower of closed string states that couple to the D-brane in
the $t$-channel (\ie terms with poles at $\alpha'\,t=\alpha'\,m^2=4n$
with $n=0,1,2,\ldots$). For low momentum transfer,
\ie $\alpha'\,t<<1$, the first term representing the exchange of
massless string states dominates. In this case, eqs.~(\ref{ampuniverse}
-- \ref{subdivide}) reduce to 
\beq
A\simeq i\,{\kappa\, T_p}\, \frac{a_1}{t}\ \ .
\labell{pole}
\eeq
One can reproduce these long-range interactions with a calculation
in the low energy effective field theory in which D$p$-brane source
terms are added to the field theory action. Alternatively, one can
think of these amplitudes as representing the interaction of the
external string states with the long range background fields
generated by the D$p$-brane (see \eg \cite{mandl}). Since ultimately
we are interested in determining these background fields, we will
make use of both interpretations of these amplitudes (\ref{pole}).

The NS-NS sector is common to both type II 
superstring theories, and so the same low energy effective action
describes the graviton, dilaton and Kalb-Ramond fields in both
theories. The latter may be written as
\beq
I_{NS-NS}=
\int d^{10}\!x\,\sqrt{-g}\,\left [\frac{1}{2\k^2}R
-\frac{1}{2}(\na \phi)^2-\frac{3}{2}\ H^2\,e^{-{\sqrt{2}\k}\phi}
\right]
\labell{nsaction}
\eeq
where $H=\frac{1}{3}(\partial_\alpha B_{\mu\nu}+\partial_\nu
B_{\alpha \mu}+\partial_\mu B_{\nu \alpha})$. 
Given this low energy effective action, one can calculate the
different propagators, interactions, and subsequently scattering amplitudes
for these three massless NS-NS particles. In doing so, one 
defines the graviton field by $g_{\mu\nu}=\eta_{\mu\nu}+2\k\,h_{\mu\nu}$.
One can verify that this action (\ref{nsaction})
correctly describes the three-point scattering
of NS-NS states on the sphere.

To consider the scattering of these particles from a D$p$-brane, we would
supplement the low energy action with source terms for the brane as
follows:
\beq
I_{source}=\int d^{10}\!x\,\left[S_B^{\mu\nu}\,B_{\mu\nu}+S_\phi\,\phi+
S_h^{\mu\nu}\,h_{\mu\nu}\right]\ \ .
\labell{souraction}
\eeq
We did not make the effort here to extend these source terms in a covariant
manner \cite{Dbract,Dbracta}, since this will be irrelevant for our 
leading order calculations. Also note that at least to leading order,
$S_B$, $S_\phi$ and $S_h$ above will be $\delta$-function sources which are
only nonvanishing at $x^i=0$ using the coordinates of the previous
section. 
\begin{figure}
\centerline{\epsfxsize 2.4 truein \epsfbox {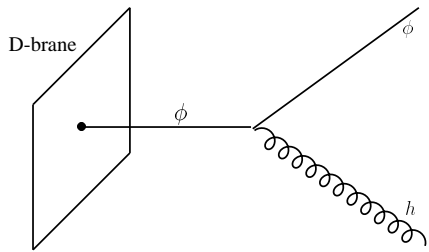}}
\caption{Feynman diagram for graviton-dilaton scattering from a D-brane }
\end{figure}
We begin by determining the dilaton source $S_\phi$. To this end, we consider
a scattering process in which an external dilaton is converted to
a graviton. The Feynman diagram corresponding
to this process appears in figure (1).
Examining the low energy action (\ref{nsaction}), one finds that
the only relevant three-point interaction is one graviton coupling
to two dilatons through the dilaton kinetic term. Thus in figure (1),
the only particle appearing in the $t$-channel is the dilaton, and 
hence this amplitude will uniquely determine $S_\phi$.
The field theory amplitude may be written
\[ 
A'_{h\phi}=i\tilde{S}_\phi(k)\,\tG_{\phi}(k^2)\,\tV_{h\phi \phi}
(\pol_1,p_1,p_2)
\] 
where $\tilde{S}_\phi(k)$ is the Fourier transform of the dilaton source,
$\tG_\phi(k^2)=-i/k^2$ is the dilaton's Feynman propagator, and
\[
\tV_{h\phi \phi}=-i\,2\k\, p_2\inn \pol_1\inn k
\]
is the vertex factor for the graviton-dilaton-dilaton interaction.
Here, $\pol_1$ is the graviton polarization tensor, and
$k^\mu=-(p_1+p_2)^\mu$ is the $t$-channel momentum. (We have
not included in $A'_{h\phi}$ a $\delta$-function which imposes
momentum conservation in the directions parallel to the D$p$-brane.)
The analogous  string amplitude $A_{h\phi}$ is constructed 
from eq.~(\ref{finone}) by inserting the appropriate
external polarization tensors from eq.~(\ref{nspol}).
Comparing $A'_{h\phi}$
with the massless $t$-channel pole in  $A_{h\phi}$,
one finds agreement by setting
\beq
\tS_\phi(k)=
-\frac{T_p}{4\sqrt{2}}\left (2+\Tr(D)\right)=-
\frac{T_p}{2\sqrt{2}}(p-3)
\labell{dilsource}
\eeq
where we used that $\Tr(D)=2p-8$.
Note that the source is a constant independent of $k$ in agreement with the
expectation that the position space source in eq.~(\ref{souraction})
is a $\delta$-function in the transverse directions.

The graviton source $S_h$ can be determined from either $h$-$h$ or $B$-$B$
scattering from the Dirichlet brane. In the first, massless
$t$-channel exchange is mediated by only a graviton, while in the
second, both a graviton and dilaton propagate in the $t$-channel.
We will consider the scattering of antisymmetric tensors here because
the relevant three-point interactions are much simpler. 
The Feynman diagrams for $B$-$B$ scattering are shown in
figure (2).
\begin{figure}
\centerline{\epsfxsize 4.7 truein \epsfbox {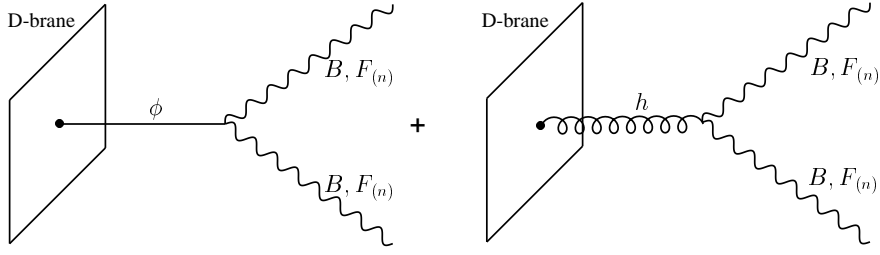}}
\caption{Feynman diagrams for scattering of two (NS-NS or R-R)
antisymmetric tensor states from a D-brane}
\end{figure}
The corresponding amplitude is 
\beq
A'_{BB}=
i\tS_h^{\mu\nu}(k)\,(\tG_h)_{\mu\nu,\lambda\rho}(k^2)\,
(\tV_{hBB})^{\lambda\rho}
+i\tS_\phi(k)\,\tG_{\phi}(k^2)\,\tV_{\phi BB}
\labell{BBamp}
\eeq
where the graviton propagator (in Feynman-like gauge ---
see \eg \cite{velt}) and the
three-point interactions are given by
\beqar
(\tG_h)_{\mu\nu,\lambda\rho}&=&-
\frac{i}{2}\left(\eta_{\mu\lambda}\eta_{\nu\rho}+
\eta_{\mu\rho}\eta_{\nu\lambda}-\frac{1}{4}\eta_{\mu\nu}
\eta_{\lambda\rho}\right)\frac{1}{k^2}
\nonumber\\
(\tV_{hBB})^{\lambda\rho}&=&-i\,2\kappa\,\left(
\frac{1}{2}\left(p_1\inn
p_2\,\eta^{\l\rho}-p_1^{\l}\,p_2^{\rho}-p_1^{\rho}\,p_2^{\l}
\right)\Tr({\pol}_1\inn{\pol}_2)\right.
\nonumber\\
&&\qquad\ -p_1\inn{\pol}_2\inn{\pol}_1\inn p_2\, \eta^{\l\rho}+
2\,p_1^{(\l}\,{\pol_2}^{\rho)}\inn\pol_1\inn p_2
+2\,{p_2}^{(\l}\,{\pol_1}^{\rho)}\inn\pol_2\inn p_1
\nonumber\\
&&\qquad\left.\ +2p_1\inn{\pol_2}^{(\l}\,{\pol_1}^{\rho)}\inn p_2
-p_1\inn p_2\,(\pol_1^{\l}\inn\pol_2^{\rho}+\pol_2^{\l}\inn\pol_1^\rho)
\vphantom{\frac{1}{2}}\right)
\nonumber\\
\tV_{\phi BB}&=&-i{\sqrt{2}\k}\left(2p_1\inn\pol_2\inn\pol_1\inn p_2
-p_1\inn p_2\ \Tr(\pol_1\inn\pol_2)\right)
\eeqar
where our notation is such that $\Tr({\pol}_1\inn{\pol}_2)
=\pol_1{}^{\mu\nu}\,\pol_{2\nu\mu}$, $p_1\inn\pol_2^\l=p_{1\delta}
\,\pol_2^{\delta\l}$ and ${}^{\rho}{\pol_1}^\rho\inn p_2=
\pol_1^{\rho\delta}\,p_{2\delta}$. Now again we must compare
this result with the massless $t$-channel pole in the
string amplitude (\ref{finone}) with an appropriate choice
of polarization tensors. 
Unravelling $\tS^{\mu\nu}$ from eq.~(\ref{BBamp}) is simplified
by noting that the only symmetric two-tensor available is 
\beq
\tS_h^{\mu\nu}(k^2)=a(k^2)\,V^{\mu\nu}
+b(k^2)\,N^{\mu\nu}+c(k^2)k^\mu k^\nu
\labell{tensor}
\eeq
given the symmetries of the scattering process. Here,
$V$ ($N$) is the metric 
in the subspace parallel (orthogonal) to the world-volume
of the D$p$-brane --- see Appendix \ref{kin}. 
Now comparing $A'_{BB}$ with the massless $t$-channel pole in
the string amplitude $A_{BB}$
fixes $a$, $b$ and $c$ to be constants with $b=c=0$ leaving
\beq
\tS_h^{\mu\nu}=-T_p\,V^{\mu\nu}
\labell{gravsource}
\eeq
This source is essentially the (Fourier transform of)
the D$p$-brane's stress energy tensor, (\ie $\tS_h^{\mu\nu}={\kappa}
\widetilde{T}^{\mu\nu})$. Hence we see that $T_p$ is essentially the
D-brane tension, and that as expected there is only stress energy in the 
world-volume directions. As a cross check, 
we have calculated $\tS_h^{\mu\nu}$ from graviton-graviton scattering, and
found the same result. One can also calculate dilaton-dilaton scattering
in which the $t$-channel interaction is mediated by a graviton, the result
is consistent with the result in eq.~(\ref{gravsource}). However,
this amplitude alone does not have enough structure to completely fix
all of the unknown functions appearing in eq.~(\ref{tensor}).
In fact, it is important in the previous calculations that we had
the fully covariant string amplitudes without any restrictions
on the polarization tensors in order completely fix the
graviton source. Finally we note that a more careful
examination of the case $p=8$ shows that the pole is cancelled
by momentum factors in the vertices for {\it all} of the amplitudes
-- see Appendix
\ref{domain}. Hence our present analysis does not determine the sources for
the eight-branes.

To determine the antisymmetric tensor source, we considered
$B$-$\phi$ and $B$-$h$ scattering. In these cases though, the 
string scattering amplitude (\ref{finone}) is easily
shown to vanish and so one
concludes that the {\it linear} source term for $B_{\mu\nu}$ 
in eq.~(\ref{souraction}) precisely vanishes.
\beq
\tS_B^{\mu\nu}=0
\labell{Bsource}
\eeq

One can also verify these results for the sources of the dilaton
and graviton fields, eqs.~(\ref{dilsource}) and (\ref{gravsource}),
by considering the scattering of the Ramond-Ramond tensor fields
from the D$p$-branes. The relevant terms in the low energy effective
action are
\beq
I=\int d^{10}\!x\,\sqrt{-g}\,\sum_n\,\left(
-\frac{8}{n!}
\,F_{(n)}\inn F_{(n)}\,e^{(5-n)\frac{\k}{\sqrt{2}}\phi}\right) \ \ .
\labell{raction}
\eeq
For the type IIa superstring, the sum runs over $n=1,$ 3 and 5,
while for the type IIb theory, the sum includes $n=2$ and 4.
An added complication is that $F_{(5)}$ should be a self-dual
field strength for which no covariant action exists.
The above non-self-dual action will yield the correct type IIb equations
of motion when the self-duality constraint is imposed by hand \cite{boon}
--- \ie one makes the substitution $F_{(5)} \rightarrow F_{(5)}+
*\!F_{(5)}$. One can verify that this action (\ref{raction}) reproduces
the three-point string amplitudes on the sphere for two R-R fields
scattering with a graviton or dilaton. For a complete description
of all of the $NS^2$-$R^2$-$R^2$ amplitudes, one would have to
include various Chern-Simon interactions between $B_{\mu\nu}$
and the R-R fields, but we will not be interested in these since
the Kalb-Ramond source vanishes, eq. (\ref{Bsource}).

The Feynman diagrams corresponding to the two-point scattering of 
R-R fields from the D$p$-branes are in figure (2).
The scattering amplitude is
\[ 
A'_{FF}=i\tS_h^{\mu\nu}\,(\tG_h)_{\mu\nu,\l\rho}(k^2)\,(\tV_{hFF})^{\l\rho}
+i\tS_\phi\,\tG_{\phi}(k^2)\,\tV_{\phi FF}
\] 
where
\beqar
(\tV_{hFF})^{\l\rho}&=&i\,\k\,
\frac{16}{n!}\left[2n\,(F_{1(n)})^{(\l}{}_{\nu_2\nu_3\cdots\nu_n}\,
(F_{2(n)})^{\rho )\nu_2\nu_3\cdots\nu_n}-
\eta^{\l\rho}\,F_{1(n)}\inn F_{2(n)}\right]
\\
V_{\phi F_{(n)}F_{(n)}}&=&
-i\,\k\frac{8\sqrt{2}}{n!}(5-n)\,F_{1(n)}\inn F_{2(n)}
\eeqar
where we have left the external momenta and polarization tensors in
the covariant form of a linearized field strength as in eq.~(\ref{strong}).
Substituting in our previous results for $\tS_\phi$ and $\tS_h^{\mu\nu}$
in eqs.~(\ref{dilsource}) and (\ref{gravsource}), $A'_{FF}$ reduces to
\beq
A'_{FF}=
i\,\k\, T_p\, \frac{8}{n!}
\frac{1}{k^2}\left(\Tr(D)\,F_{1(n)}\inn
F_{2(n)}-2n\,D^{\l}{}_{\rho}\,(F_1)_{\l\nu_2\nu_3\cdots\nu_n}\,
(F_2)^{\rho\nu_2\nu_3\cdots\nu_n}\right)\ \ .
\labell{FFpole}
\eeq
To demonstrate that this result agrees with the string amplitude,
we focus on the $a_1$ contribution in the kinematic factor
(\ref{kinrrrr})
\beqar
a_1&=&-\frac{1}{2}\Tr(P_-\slf_{1(n)}M\g_\mu C^{-1} M^T\slf^T_{2(m)}C\g^\mu)
\\
&=&-\frac{1}{2}Tr\left(P_-\Gamma_{1(n)}\,\g^{\mu}\,\Gamma_{2(m)}\,\g^{\nu}\right)
\,D_{\mu\nu}\,(-1)^{\frac{1}{2}m(m+1)}
\\
&=& -\frac{8}{n!}\,\delta_{mn}\left [\Tr(D)F_{1(n)}\inn F_{2(n)}-
2n\,D^{\l}{}_{\k}\,F_{1\,\l\nu_2
\nu_3\cdots\nu_n}\,{F_2}^{\k\nu_2\nu_3\cdots\nu_n}\right]
\eeqar
where we have applied various spinor matrix identities from
Appendix \ref{spinor}. Combining this result with eq.~(\ref{pole}),
we find that the result coincides precisely with eq.~(\ref{FFpole}).
As an aside, we also note that this explicit evaluation of $a_1$
reproduces the result of Ref.~\cite{igorb}. The agreement between the
field theory and string scattering of R-R particles was also found
in the latter reference.

{}From the dilaton and graviton sources, it is a simple matter to
calculate (the Fourier transform of) the corresponding long range
fields around the D$p$-branes. These fields are precisely the product
of the source and the Feynman propagator in the transverse momentum
space. Hence the long range dilaton field is
\beqa
\tilde{\phi}(k^2)&=&i\tS_\phi\,\tG_{\phi}(k^2)=
-\frac{T_p}{4\sqrt{2}}\,\frac{2+\Tr(D)}{k^2}
\nonumber\\
&=&-\frac{T_p}{2\sqrt{2}}\,\frac{p-3}{k^2}\ \ .
\labell{dilfield}
\eeqa
Similarly the long range gravitational field becomes
\beqa
\tilde{h}_{\mu\nu}(k^2)&=&i\tS_h^{\l\rho}\,(\tG_h)_{\l\rho,\mu\nu}
\nonumber\\
&=&-\frac{T_p}{8k^2}\left((7-p)\,V_{\mu\nu}-(p+1)\,N_{\mu\nu}\right)
\labell{gravfield}
\eeqa
Of course given eq.~(\ref{Bsource}), the long range antisymmetric 
tensor field vanishes.

\begin{figure}
\centerline{\epsfxsize 2.4 truein \epsfbox {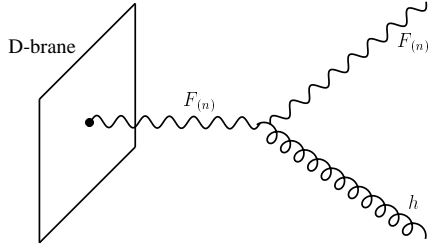}}
\caption{Feynman diagram for graviton--R-R tensor scattering from a D-brane}
\end{figure}
We would also like to determine the long range Ramond-Ramond fields
around the D$p$-brane. In this case, introducing a local source for a
``magnetic'' charge is a slight complication, which we avoid by working
directly with the background fields. In this case, one expands
the action (\ref{raction}) around a background field strength
(which satisfies the equations of motion). The relevant
amplitudes arise from three-point amplitudes involving one
background field and two external particles. Essentially as in
eqs.~(\ref{dilfield}) and (\ref{gravfield}), one is replacing
the source and propagator in the previous calculations by the
background field in the amplitude. Here we consider a scattering
process in which a graviton converts to a R-R tensor field.
The Feynman diagram is shown in figure (3), and
the corresponding scattering amplitude is
\[ 
A'_{hF}=i\,\k\,n\,\frac{32}{n!}\,\pol_{2}^{\l\mu}
\,(F_{1(n)})_\l{}^{\nu_2\cdots\nu_n}\,
(\tilde{F}_{(n)})_{\mu\nu_2\cdots\nu_n}
\]
where $\pol_2$, $F_{1(n)}$ and $\tilde{F}_{(n)}$ are the graviton
polarization, the external R-R particle's (linearized) field strength,
and the background R-R field strength, respectively. Note that the interaction
requires that the form degree $n$ is the same for both the external
and the background R-R fields. 
The massless $t$-channel pole for the string scattering amplitude
$A_{hF}$ comes from $a_1$ in eq.~(\ref{kinrrnn}) which yields
\[ 
a_1=\pm
\,i\,\frac{8\sqrt{2}}{(n-1)!}\,\pol_{2}^{\l\mu}\,(F_{1(n)}
)_\l{}^{\nu_2\cdots\nu_n}\times\left\lbrace\matrix{
nk_{[\nu_n}\,(\e^v)_{\mu\nu_2\cdots\nu_{n-1}]}
&\qquad{\rm for}\ p=n-2\cr
k^{\rho}\,(\e^n)_{\rho\mu\nu_2\cdots\nu_n}
&\qquad{\rm for}\ p=8-n\ \ .\cr}
\right.
\] 
where the $\pm$ sign is the same as that appearing in choice
made for $M$ from eq.~(\ref{finalm}).
Comparing the string amplitude $A_{hF}$ with $A'_{hF}$, one obtains
\[ 
\tilde{F}_{\nu_1\nu_2\cdots\nu_n}=\mp i\frac{T_p}{2\sqrt{2}k^2}\times
\left\lbrace\matrix{
nk_{[\nu_n}\,(\e^v)_{\nu_1\cdots\nu_{n-1}]}&\qquad{\rm for}\ p=n-2\cr
k^\mu\,(\e^n)_{\mu\nu_1\nu_2\cdots\nu_n}&\qquad{\rm for}\ p=8-n\ \ .\cr}
\right.
\] 
Anticipating that the first line corresponds to an ``electric" field, we see
D$p$-branes with $p=-1$, 0, 1, 2, and 3 have ``electric" fields
with $n=1,$ 2, 3, 4 and 5, respectively. While the second line above shows
that D$p$-branes with $p=3,$ 4, 5, 6 and 7 have ``magnetic'' fields
with $n=5$, 4, 3, 2 and 1, respectively. As expected the
D3-brane simultaneously carries electric and magnetic fields from
the self-dual five-form.

\subsection{Comparison of fields} \labels{compare}

To make a precise comparison of the results in the previous two
sections, we must first take into account that the low energy string actions,
(\ref{nsaction}) and (\ref{raction}), have a different normalization
from that used to derive the $p$-brane solutions, eq.~(\ref{actors}).
Hence we make use the
following field redefinitions for the fields appearing in section
\ref{lowsol}
\beqa
\hat{h}_{\mu\nu}&\equiv&2\k\, {h'}_{\mu\nu}
\nonumber\\
\sqrt{\hat{\g}}\hphi&\equiv&{\sqrt{2}\k}\,\phi'
\labell{redefine}\\
\hat{F}_{(n)}&\equiv&4\sqrt{2}\k \,F'_{(n)}
\nonumber
\eeqa
and we identify the constants $\sqrt{\hat{\g}}\ha\equiv a'$
and $8\pi\hat{G}_N=\k^2$, as well as setting $\hD=10$.
With these choices the $p$-brane action (\ref{actors}) becomes
\[ 
\hat{I}= 
\int d^{10}\!x\,\sqrt{-g'}\left [\frac{1}{2\k^2}R'(g')-
{1\over 2}(\nabla \phi')^2
-{8\over n!}e^{-a'{\sqrt{2}\k} \phi'}F^{\prime 2}_{(n)} \right]\ \ .
\] 
where $g'_{\mu\nu}=\eta_{\mu\nu}+2\k\, {h'}_{\mu\nu}$.
Finally this action my be identified with the relevant part of
the effective superstring action by setting $a'=(n-5)/2$.

To compare the low energy  p-brane solutions to the 
long-range D$p$-brane fields, we make a Fourier transform 
of the asymptotic fields of sect.~\ref{lowsol}. The essential
transform is that for $G(|\vec{x}|/\ell)$ in eq.~(\ref{green})
which is the Green's function in the transverse space.
\beq
\tG(\vec{k}^2)=\frac{\cA_{\hd-1}\,\ell^{\hd-2}}{\vec{k}^2}
\labell{greener}
\eeq
where $\vec{k}$ is a wave-vector in the subspace orthogonal
to the $p$-brane, and as before $\cA_{\hd-1}$ %=2\pi^{\hd}/\Gamma(\hd/2)$
is the area of a unit $(\hd-1)$-sphere.
The Green's function (\ref{greener}) has the same form as the Feynman
propagators appearing in the field theory calculations with $k^2=\vec{k}^2$
--- recall that the $t$-channel momentum vector
is a spatial vector in the transverse subspace
(see Appendix \ref{kin}). 

Now in terms of the primed fields introduced in eq.~(\ref{redefine}),
the Fourier transform of the asymptotic fields
(\ref{asymetric}--\ref{elecfields}) of the
$p$-brane solutions with ``electric'' charge are
\beqar
{\tilde{h}'}_{\mu\nu}&\simeq&\frac{\cA_{\hd-1}\mu\ell^{\hd-2}}{\k}
\frac{1}{\vec{k}^2}
{\rm diag}(-\alpha , \alpha ,\ldots,\alpha , \beta ,\ldots\beta)\\
\tilde{\phi}'&\simeq&-\frac{\ha\sigma\cA_{\hd-1}\mu\ell^{\hd-2}}{2\sqrt{2}\k}
\frac{1}{\vec{k}^2}\\
\tilde{F}'_{(n)}&\simeq&\mp i\frac{\sqrt{\sigma}\cA_{\hd-1}\mu\ell^{\hd-2}}
{4\k}\frac{1}{\vec{k}^2}\,\vec{k}\inn dx\wedge (\e^v)\ \ .
\eeqar
In this case with $d=10$, $\hd=9-p$ and $p=n-2\le 3$,
the constants (\ref{exponents}) reduce to
\beqa
\alpha&=-\frac{7-p}{16} 
\qquad\beta&=\frac{p+1}{16}
\nonumber\\
\sigma&=\frac{1}{2}\qquad\hfill\ha&=n-5=p-3\ \ .
\labell{exponentelec}
\eeqa
Hence we have complete agreement between the long-range D$p$-brane
fields and those above if $\kappa T_p=\cA_{8-p}\mu\ell^{7-p}/2$.
Further applying the mass formula (\ref{mess}), we find the
mass per unit $p$-volume is given by $M_p=T_p/\k$, while the 
electric charge
per unit $p$-volume (\ref{elecharge}) is $Q_p=\pm \sqrt{2} T_p$.

For the magnetically charged $p$-branes, the metric is unchanged
while the Fourier transform of eq.~(\ref{magfields}) yields
\beqar
\tilde{\phi}'&\simeq&+\frac{\ha\sigma\cA_{\hd-1}\mu\ell^{\hd-2}}{2\sqrt{2}\k}
\frac{1}{\vec{k}^2}\\
{\tilde{F}}'_{(n)}&\simeq&\mp i
\frac{\sqrt{\sigma}\cA_{\hd-1}\mu\ell^{\hd-2}}{4\k}
\frac{1}{\vec{k}^2}\,i_{\vec{k}}(\e^n)\ \ .
\eeqar
Again with $d=10$, $\hd=9-p$ and now $p=8-n\ge 3$, one finds the
same expressions for the constants $\alpha$, $\beta$ and $\sigma$ as
in (\ref{exponentelec}), while
\[ 
\ha=n-5=3-p
\] 
has the opposite sign.
Again with $\kappa T_p=\cA_{8-p}\mu\ell^{7-p}/2$, one finds
precise agreement between the fields of the D$p$-branes and extremal p-brane
solutions.
The magnetic charge per unit $p$-volume (\ref{magcharge}) is
$Q_p=\pm \sqrt{2}T_p$. 

\section{Discussion} \labels{discuss}

In this paper, we have presented detailed calculations of all
two-point amplitudes describing massless closed type II superstrings
scattering from a
Dirichlet $p$-brane in ten dimensions. Using these results we derived the long
range fields around D$p$-branes for $0\le p\le 7$, and found that
they correspond to those of the low energy solutions
describing a supersymmetric $p$-brane carrying a R-R charge.

One of the most interesting aspects of this work was the observation that
these closed string two-point amplitudes describing scattering from
a D$p$-brane are simply related to four-point amplitudes for type I
open superstrings. This relation allowed the D-brane amplitudes to
be easily determined using the previously calculated open
string amplitudes. This result is similar to the work of
Kawai, Lewellen and Tye \cite{klt}, who were able to express
tree-level closed string amplitudes as products of pairs of open string
amplitudes along with certain ``sewing'' factors. This structure arises
because of the independence of the correlation functions for the
left- and right-moving sectors. The present situation is similar
except that the D-brane boundary naturally ``sews'' 
the right- and left-movers together in a single
open string amplitude. In Ref.~\cite{klt}, the results for
tree-level string scattering are extended to amplitudes with an arbitrary
number of closed string states. It would be interesting to find
a similar extension relating a D-brane scattering amplitude
for $N$ closed strings to a Type I amplitude for $2N$ open
strings.

Our normalizations are chosen such that $T_p$ coincides precisely
with the string tension in a standard analysis, \eg \cite{Dbreview}.
The physical tension \cite{Dbreview} is precisely the mass 
per unit $p$-volume $\tau_p=T_p/\k=M_p$. 
For a fundamental D$p$-brane then,
$T_p=\sqrt{\pi}/(2\pi\sqrt{\alpha'})^{p-3}$.
Higher multiples of this fundamental value of $T_p$ would arise for
D$p$-branes which are superpositions or bound states
of fundamental branes.
Similarly our charges correspond precisely to
those given in the standard analysis \cite{joe,Dbreview},
\ie $\mu_p=Q_p$ with
$Q_p^2=2T_p^2$. The sign of the charges is undetermined.
It is fixed in the conformal field theory calculations by making
a choice of sign for the matrix $M$ appearing in the spin operators
--- see Appendix \ref{spinor}.
It is remarkable that the conformal field theory allows just
enough ambiguity in $M$ to accommodate both branes and anti-branes.

This matrix also plays an important role in the spacetime supersymmetry of the
D-brane amplitudes, as follows. On a closed type II superstring
world sheet, one can construct two independent spacetime supersymmetry
currents \cite{fms,danf}
\[
Q_L=\e_L^A\,e^{-\phi(z)/2}\,S_A(z)\qquad\quad{\rm and}
\qquad\quad Q_R=\e_R^A\,e^{-\tilde{\phi}(\bz)/2}\,\tS_A(\bz)
\]
with independent actions on the left- and right-moving modes on
the world sheet. Hence $\e_L$ and $\e_R$ are independent
Majorana-Weyl spacetime supersymmetry parameters in the Type II
theory. In the Type I
open superstring theory, the left- and right-movers are tied together
by the world sheet boundaries, and hence there is a single SUSY current
to act on the open string states. Given the relation
of open string and D-brane scattering amplitudes, the latter
would inherit this single spacetime supersymmetry of the open
superstring amplitudes. Hence there will be
a single SUSY current which acts
consistently on both the left- and right-moving components of the
closed string vertex operators. If we extend the spin fields in the SUSY
currents to the entire complex plane using eq.~(\ref{replaced})
then the right-moving current becomes
\[
Q_R=\e_R^B\,M_B{}^A\,e^{-{\phi}(\bz)/2}\,S_A(\bz)\ \ .
\]
In order to construct the single consistent current then we must set 
\[
\e_L^A=\e_R^B\,M_B{}^A\ \ .
\]
This then is a conformal field theory approach to understanding that
the D-branes preserve precisely one half of the spacetime
supersymmetries of the Type II theory\cite{joe}. 

The long range fields around the D$p$-branes were determined in
sect.~\ref{back} by considering the massless $t$-channel poles
in the string scattering amplitudes. As well as these poles, the
string amplitudes contain an infinite number of massive poles as
well. Using a standard expansion of the Euler beta function
(see \eg \cite{gsw}), one finds that eq.~(\ref{ampuniverse})
may be written as
\beq
A=-i\,\k\,T_p\left(\frac{a_1}{k^2}-\frac{a_2+2q^2a_1}{k^2+2}
+\frac{(2q^2-1)(a_2+2q^2a_1)}{k^2+4}+\cdots\right)
\labell{tchannel}
\eeq
where $k^2=-t$. Each of the higher $t$-channel poles represents
a massive closed string state coupling to the D$p$-brane. From this
point of view, the D$p$-brane provides a $\delta$-function source for
each of these massive states, just as it does for the massless states.
The same set of $\delta$-functions in the transverse coordinates appears
in the construction of a boundary state description of a D-brane
(see \eg \cite{igorc}). This would then lend itself to an interpretation
of D-branes as objects of zero thickness. However, since as seen above
a D-brane is not
only a source of the massless fields but also massive fields with
$m^2=4n/\alpha'$, the conventional (low energy) spacetime picture will
breakdown at distances of the order of $\sqrt{\alpha'}$. It is within this
range that the full closed string spectrum makes its presence felt.
Hence from this perspective, one would ascribe a thickness of the
order of $\sqrt{\alpha'}$ to D-branes.

Of course,
the preceding discussion ignores the stringy nature of the amplitudes
and in particular their $s$-$t$ channel duality. As emphasized by
\cite{igora}, the amplitudes can be reorganized as a series of
$s$-channel (or $q^2$) poles
\beq
A=i\,\frac{\k\,T_p}{2}\left(\frac{a_2}{2q^2}+\frac{a_1+k^2a_2/2}{2q^2+1}
+\frac{(1-k^2/2)(a_1+k^2a_2/2)}{2q^2+2}+\cdots\right)\ \ .
\labell{schannel}
\eeq
In this case, the poles coincide with open string states moving along
the D-brane. These in turn correspond to excitations of internal modes
or deformations of the D-brane. Hence this point of view naturally
displays a smearing of the D-brane again on the order of $\sqrt{\alpha'}$
\cite{igora}.

An exception to the above form of the amplitudes is the special case of
$p=-1$. The D(--1)-brane corresponds to a Euclidean instanton which then
has no world volume and hence there is no momentum flow parallel to the
world volume --- \ie $q^\mu=0$. As noted in \cite{igora}, the amplitudes
can then not take the form in eq.~(\ref{schannel}) and further without
$s$-$t$ channel duality one would expect that the infinite series
in the $t$-channel expression (\ref{tchannel}) is also truncated.
Explicitly calculating the amplitudes considered in the present
investigation (\ie for massless external states) shows that in fact
only the massless
$t$-channel pole survives. This was already observed in Ref.~\cite{igora}
in the scattering amplitude for two antisymmetric tensors
\beq
A_{BB}=i\k T_p\left(\frac{4\,p_1\inn\pol_2\inn\pol_1\inn p_2}{t}- 
Tr(\pol_1\inn\pol_2)\right)\ \ .
\labell{bbscat}
\eeq
(Also the graviton-graviton amplitude vanishes completely \cite{igora}.)
A similar result holds for graviton-dilaton scattering and
graviton--R-R tensor scattering
\beqa
A_{h\phi}&=&-i2\sqrt{2}\,\k\,T_p\frac{p_2\inn\pol_1\inn p_2}{t}
\nonumber\\
A_{h F}&=&-8\sqrt{2}\k\,T_p\frac{\pol_{1}^{\l\mu}\,(F_2)_\l k_\mu}{t}\ \ .
\labell{otscat}
\eeqa
These amplitudes were determined by substituting
$D_{\mu\nu}=-\eta_{\mu\nu}$ into our general results and taking
the limit that $q^2\rightarrow0$. Recalculating these amplitudes
directly from conformal field theory reproduces the same answers.
Given that these three massless poles are the same as in our general formulae,
the analysis of sect.~\ref{back} applies to D(--1)-branes as well.
Hence their long range fields are those of the low energy
solutions discussed in Ref.~\cite{instanton} (see also \cite{igorb}).

While no massive poles in these particular amplitudes
(\ref{bbscat}-\ref{otscat}), one should
not conclude that no massive string states couple to the D(--1)-branes.
We expect that such poles and couplings would make their appearance
in scattering amplitudes involving massive external strings. This
should be evident given the complex form of the boundary state
describing a D(--1)-brane \cite{bound}. Hence even for these
instantons, one would ascribe a thickness of the order of $\sqrt{\alpha'}$.

%Note that the Kalb-Ramond scattering
%amplitude (\ref{bbscat}) satisfies the Ward identity corresponding
%to one-form gauge invariance
%-- \eg $A_{BB}$ vanishes upon $\pol_{1\mu\nu}=p_{1\mu}\l_{1\nu}-
%p_{1\nu}\l_{1\mu}$. On the other hand, neither of the graviton
%amplitudes in (\ref{otscat}) satisfies the diffeomorphism
%Ward identity. In the conformal field theory 

Our analysis of the long range fields is not valid in the case
of domain walls (for which $p=8$) -- see Appendix \ref{domain}.
Naively the amplitudes for two NS-NS or two R-R states appear to
give the expected long range gravitational and dilaton fields.
However, a closer inspection of these amplitudes shows that because of
the unusual kinematics of this configuration, there is
no pole at $t=0$. This result is in fact in agreement with the analogous
field theory calculations. In order to overcome this difficulty,
one would have to begin by extending the source action (\ref{souraction})
to include nonlinear terms. These interactions are required to understand
the two-particle contact terms that arise in the string amplitudes.
It should be possible to extract the necessary terms from the
covariant Born-Infeld actions constructed as the effective
D-brane actions \cite{Dbract,Dbracta}. Another necessary ingredient
here would be a better understanding of the vertex operator description
of the $n=10$ R-R field strength whose charge is carried
by the D8-branes.

\section*{Acknowledgments}
We gratefully acknowledge useful conversations with Clifford Johnson, Vipul
Periwal and Joe Polchinski. We also thank
Omid Hamidi-Ravari for helping us to generate the figures.
This research was supported by NSERC of Canada and Fonds FCAR du
Qu\'ebec. M.R.G. was supported by Ministry of Culture and Higher
Education of Iran.

\appendix

\section{Kinematics of Two-Point Amplitudes} \labels{kin}

In the calculation of Dirichlet 
$p$-brane scattering amplitudes, it is convenient
to introduce two sets of frame fields: $v^a{}_\mu$ with $a=0,\ldots,
p$, and $n^a{}_\mu$ with $a=p+1,\ldots,9$. So the $v^a$ are unit vectors
tangent to the $p$-brane's world-volume while the $n^a$ are orthogonal
the world-volume. The vector (Greek) indices are raised and lowered as
usual with the metric $\eta_{\mu\nu}=diag(-1,+1,\ldots,+1)$, 
and similarly the frame (Latin)
indices are raised and lowered with $\eta_{ab}$. Then one has
\[
n^a\inn n^b=\eta^{ab} \qquad v^a\inn v^b=\eta^{ab}\qquad
n^a\inn v^b=0\ .
\]
Recall that it is the coordinate fields running orthogonal to
the D$p$-brane and hence parallel to the
$n^a{}_\mu$, which satisfy the Dirichlet boundary conditions.

Given this separation of the frame fields, one can construct two useful
projection operators, namely,
\[
N_{\mu\nu}=n^a{}_\mu n_{a\nu}\qquad{\rm and}
\qquad V_{\mu\nu}=v^a{}_\mu v_{a\nu}
\]
where $N$ projects vectors into the transverse subspace or the
subspace orthogonal to the D$p$-brane,
and $V$
projects vectors into the subspace parallel to the D$p$-brane.
Hence $N_\mu{}^\lambda N_{\lambda\nu}=N_{\mu\nu}$,
$V_\mu{}^\lambda V_{\lambda\nu}=V_{\mu\nu}$ and
$N_\mu{}^\lambda V_{\lambda\nu}=0$.
Completeness of the basis of frames also yields 
\[
\eta_{\mu\nu}=V_{\mu\nu}+N_{\mu\nu}\ .
\]
The matrix appearing in the correlation functions for
coordinate fields (\ref{xcor}) and world-sheet spinors (\ref{psicor})
is given by
\beq
D_{\mu\nu}=V_{\mu\nu}-N_{\mu\nu}\ .
\labell{dmatrix}
\eeq
Combining the above expressions, one also has
$D_{\mu\nu}=\eta_{\mu\nu}-2N_{\mu\nu}=2V_{\mu\nu}-\eta_{\mu\nu}$.
Also note that $D_\mu{}^\lambda D_{\lambda\nu}=\eta_{\mu\nu}$.

These frames are also useful in defining certain volume forms.
In the subspace parallel to the brane's world-volume,
one has a $(p+1)$-form
\[
\ev_{\mu_0\cdots\mu_p}=(p+1)!\,v^0{}_{[\mu_0}\ldots v^p{}_{\mu_p]}\ \ ,
\]
while in the normal subspace one can define the $(9-p)$-form
\[
\en_{\mu_{p+1}\cdots\mu_9}=(9-p)!\,n^{p+1}{}_{[\mu_{p+1}}\ldots
n^9{}_{\mu_9]}\ \ .
\]
The antisymmetrization of the indices above is normalized such
that both $\e^v$ and $\e^n$ take values $\pm 1$ and 0. These two
forms are related by the identities
\beqar
\ev_{\mu_0\cdots\mu_p}=\frac{1}{(9-p)!}
\e_{\mu_0\cdots\mu_p\mu_{p+1}\cdots\mu_9}\,\en^{\mu_{p+1}\cdots\mu_9}
\\
\en_{\mu_{p+1}\cdots\mu_9}=-\frac{1}{(p+1)!}
\ev^{\mu_0\cdots\mu_p}\,\e_{\mu_0\cdots\mu_p\mu_{p+1}\cdots\mu_9}
\ \ .
\eeqar
where $\e$ is the regular volume form in the full ten-dimensional
Minkowski space. As forms, one also has $\e=\ev\wedge\en$.

To describe the kinematics of the string scattering amplitudes
in a D-brane background, we divide the momentum vectors into
their parallel and transverse components, \ie
$p^\mu=(N\inn p)^\mu+(V\inn p)^\mu$. Now
one has only momentum conservation in the parallel subspace, \ie
$\sum_i (V\inn p_i)^\mu =0$, so for the two point amplitudes considered in
the paper, we may define 
\[
(V\inn p_1)^\mu=q^\mu=-(V\inn p_2)^\mu\ .
\]
Further since we are considering  massless external
states, one finds
\beq
(N\inn p_1)^2=-q^2=(N\inn p_2)^2\ .
\labell{normp}
\eeq
So $q^\mu$, the momentum flowing through the scattering amplitude
parallel to the $p$-brane, must be time-like since by definition
$(N\inn p)^\mu$ is space-like. 

As well as $p_1$ and $p_2$, $D\inn p_1$ and $D\inn p_2$ appear as momenta
appear in the vertex operators in sects.~\ref{nsns} and \ref{rest}.
Some useful identities satisfied by these vectors are
\beqar
(p_1+ D\inn p_1+p_2+D\inn p_2)^\mu=0\\
p_1\inn D\inn D\inn p_1= p_1^2=0\\
p_2\inn D\inn D\inn p_2= p_2^2=0\\
p_1\inn D\inn p_1=p_2\inn D\inn p_2=2q^2 \\
p_1\inn D\inn D\inn p_2=p_1\inn p_2=-t/2\\
p_1\inn D\inn p_2=-2q^2+t/2
\eeqar
where $t=-(p_1+p_2)^2$ is the momentum transfer to the D$p$-brane.

\subsection{Domain Wall Kinematics} \labels{domain}

The kinematics of our two-point amplitudes is special for the case
of $p=8$. In this case, the $p$-brane forms a domain wall dividing the
ten-dimensional spacetime  into two halves. The transverse space is
only one-dimensional running along $x^9$, and so it is only the ninth
component of the momentum vectors which is not conserved. Hence
eq.~(\ref{normp}) becomes
\[
(p_1^9)^2=-q^2=(p_2^9)^2\ \ .
\]
Since there is a single component of momentum appearing here, one
has either $p_1^9=-p_2^9$ or $p_1^9=p_2^9$. In the
first case, there is no momentum transfer to the brane, \ie the momentum
is precisely conserved, and hence there is no string scattering.
We will not consider this case further.
In second case, one has 
\[
t=-(p_1+p_2)^2=-(p_1^9+p_2^9)^2=-4(p_1^9)^2=4q^2\ \ ,
\]
and hence the momentum transfer is directly proportional to the
momentum flowing parallel to the brane. This result allows one to
rewrite the universal form of the amplitudes (\ref{ampuniverse}) as
\beqar
A(1,\,2)&=&-i\,\frac{\kappa\,T_p}{2}\,
\frac{\G(-2q^2)\G(2q^2)}{\G(1)} K(1,\,2)\\
&=&i\,\frac{\kappa\,T_p}{4q^2}\,\frac{\pi}{\sin(2\pi q^2)} K(1,\,2)\ \ .
\eeqar
This expression combines all of the poles in the single sine function
factor. One must remember though that
the poles for $q^2<0$ are associated with the $s$-channel,
while those with $q^2>0$ are $t$-channel poles. Naively there
is a problem at $q^2=0$ where one might expect a pole in both
channels. In fact, the apparent double pole 
can be seen to be reduced to a simple pole by the fact
that all of the terms in any of the kinematic factors $K$ always 
contain factors of $q^2$ or $t=4q^2$. However, a closer examination shows that
in fact there is no pole at all. For example consider the factors
appearing in $K_{NS-NS,NS-NS}$ in eqs.~(\ref{fintwo}) and (\ref{finthree}).
It is straightforward to show that all of the nonvanishing inner products of
momenta with polarization tensors also yield factors of $p_1^9$, \eg
\beqar
\pol_{1\mu\nu}\,p_2^\nu&=&\pol_{1\mu\nu}\,V^{\nu\rho}\,p_{2\rho}
+\pol_{1\mu\nu}\, N^{\nu\rho}\, p_{2\rho} \nonumber\\
&=&-\pol_{1\mu\nu}\, V^{\nu\rho}\, p_{1\rho}+\pol_{1\mu\nu}\, N^{\nu\rho} 
\,p_{1\rho} \nonumber\\
&=&-\pol_{1\mu\nu}\, p_1^\nu+2\,\pol_{1\mu\nu}\, N^{\nu\rho}\, p_{1\rho}
\nonumber\\
&=&2\,\pol_{1\mu\nu}\, N^{\nu\rho}\,p_{1\rho}=2\,\pol_{1\mu9}\,p_1^9 \ \ ,
\eeqar
where $N^{\nu\rho}\, p_{1\rho} =\delta^\nu_9\, p_1^9$. 
Similarly many of the terms also vanish through identities
such as $\pol_1\inn D\inn p_2=0$. In any event, all of the
remaining terms carry a factor of $(p_1^9)^2=-q^2$ which cancels
even the remaining pole.

The absence of a massless $t$-channel pole is the reason
that we make no detailed comparison of the string and field theory
scattering for eight-branes in sect.~\ref{back}. This unusual 
kinematics applies equally well for the field theory
calculations as the string amplitudes. In both cases then,
these $t$-channel terms make contributions equivalent to two-particle
contact terms on the D$p$-brane. In the string amplitude, one also
expects to find such contributions from explicit contact interactions
and also from the remnants of the massless $s$-channel pole.
In fact there are a large number of cancellations amongst these terms
in the string amplitudes. For example,
the NS$^2$-NS$^2$ amplitude in eq.~(\ref{finone}) reduces to
\beq
A=i\,\frac{\kappa\,T_p}{2}\,\Tr(\pol_1\inn D\inn\pol_2\inn D) + O(q^2)
\labell{contact}
\eeq
for small $q^2$. The appearance of two $D_{\mu\nu}$ in this term
suggests that its origin is in an explicit contact
interaction as illustrated in figure (4).
In the field theory, one expects that the covariant extension of
the source actions \cite{Dbract,Dbracta} would also naturally introduce
multiparticle contact terms to our source actions (\ref{souraction}).
Hence our present
field theory calculations can not be expected to account 
for a term such as that in eq.~(\ref{contact}). 
Introducing a fully covariant source action would be one
step towards overcoming these obstructions
to determining the eightbrane background fields, but one would
still have to unravel the $s$-channel contributions in the
string amplitude.
\begin{figure}
\centerline{\epsfxsize 2.0 truein \epsfbox {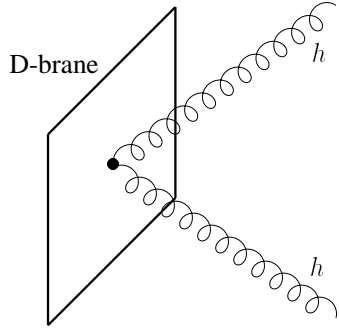}}
\caption{Feynman diagram for a two-graviton contact interaction
on a D-brane}
\end{figure}

\section{Spinors and Dirac Matrices} \labels{spinor}

We have adapted our spinor conventions within conformal field theory from
Refs.~\cite{danf,pkllsw}. We distinguish spinor and
adjoint spinor indices as (upper case Latin) subscripts and
superscripts, respectively \cite{danf}. The charge conjugation matrix
acts as a metric for raising or lowering the spinor indices, \eg
\beq
u^A=C^{AB}\,u_B\qquad\qquad u_A=C^{-1}_{AB}\,u^B\ \ .
\labell{conjugate}
\eeq
Of course, we also have $C^{-1}_{AB}\,C^{BC}=\delta^C_A$.
Following \cite{danf}, we adopt the convention that
$C^{AB}=-C^{BA}$. The spinors appearing in the open string
amplitudes (\ref{opentot}-\ref{openthree}) are Majorana, and
hence in our notation, $\bar{u}^A=u_B\,C^{BA}=-u^A$.

For our Dirac matrix conventions, we begin with the anticommutator
\[
\{\g^{\mu},\g^{\nu}\}=-2\,\eta^{\mu\nu}\ \ .
\]
Unless otherwise indicated,
implicitly in the above expression and
the various products of sect.~\ref{rest}, 
we understand that the indices on the Dirac matrices 
appear as $(\g^\mu)_A{}^B$. Using eq.~(\ref{conjugate}),
we write
\[
(\g^{\mu})_{AB}=C^{-1}_{BC}\,(\g^{\mu})_A{}^C
\qquad\qquad
(\g^{\mu})^{AB}=C^{AC}\,(\g^{\mu})_C{}^B\ \ .
\]
Then we chose a representation of the Dirac
matrices in which \cite{danf}
\[
(\g^{\mu})_{AB}=(\g^{\mu})_{BA}\ \ .
\]
In our notation, we write the transpose matrices as
\[
(\g^T_\mu)^B{}_A=(\g_\mu)_A{}^B
\]
and with the above conventions, we have
\[ 
C^{AC}\,(\g_\mu)_C{}^D\,C^{-1}_{DB}=-(\g^T_\mu)^A{}_B
\quad{\rm or}\quad 
(\g_\mu)_A{}^B=-C^{-1}_{AC}\,(\g^T_\mu)^C{}_D\,C^{DB}\ .
\]
It is also useful to note the general result that
$R_{AB}\,S^{BC}=-R_A{}^B\,S_B{}^C$ where $R$ and $S$ are
arbitrary products of $\g$-matrices.

We also have
\[
\g_{11}=
\frac{1}{10!}\,\e_{\mu_0\cdots\mu_9}\,\g^{\mu_0}\cdots\g^{\mu_9}
=\g^0\,\g^1\cdots\g^9
\]
which satisfies
\[
(\g_{11})^2=1\ ,\ \qquad
(\g_{11})_{AB}=(\g_{11})_{BA}\ ,\ \qquad
\g_{11}{}^T=-C\,\g_{11}\,C^{-1}\ .
\]
With $\g_{11}$ one constructs the chiral projection operators,
$P_{\pm}=\frac{1}{2}(1\pm\g_{11})$. Two useful identities are
\[
(P_{\pm})_{AB}=-(P_{\mp})_{BA}\qquad\quad P_{\pm}{}^T=C\,P_{\mp}\,C^{-1}\ .
\]
In constructing the R-R vertex operators, one also encounters the
matrices $\slf_{(n)}$ which are defined in eq.~(\ref{self}) as
\[
\slf_{(n)}=\frac{a_n}{n!}F_{\mu_1\cdots\mu_n}\,\g^{\mu_1}\cdots\g^{\mu_n}
\ \ .
\]
These can be shown to satisfy
\[
(\Gamma_{(n)})_{AB}=-(-1)^{\frac{1}{2}n(n+1)}(\Gamma_{(n)})_{BA}
\qquad\quad
\Gamma_{(n)}{}^T=(-1)^{\frac{1}{2}n(n+1)}C\,\Gamma_{(n)}\,C^{-1}\ .
\]
Given the above, one can show that $P_-\,\Gamma_{(5)}=0$ if the
corresponding field strength is antiselfdual, \ie $F_{(5)}=-*\!F_{(5)}$,
while $P_-\,\Gamma_{(5)}=\Gamma_{(5)}$ for a selfdual field strength,
\ie $F_{(5)}=*\!F_{(5)}$.

Finally, we would like to fix the matrix $M$
required in eq.~(\ref{replaced}) 
\beq
\tS_A(\bz)\rightarrow M_A{}^B\, S_B(\bz)
\labell{replacedagain}
\eeq
to extend the spin fields to the entire complex plane.
We begin by considering the the following operator products
\cite{pkllsw}
\beqar
\psi^{\mu}(z)\,S_A(w)&\sim&(z-w)^{-\frac{1}{2}}\,
\frac{1}{\sqrt{2}}(\g^{\mu})_A{}^B\,S_B(w)
+\ldots
\\
\tilde{\psi}^{\mu}(\bz)\,\tS_A(\bw)&\sim&(\bz-\bw)^{-\frac{1}{2}}\,
\frac{1}{\sqrt{2}}(\g^{\mu}
)_A{}^B\, \tilde{S}_B(\bw)+\ldots
\eeqar
Making the replacement in eq.~(\ref{replacedagain}) as well as
$\tilde{\psi}^{\mu}\rightarrow D^{\mu}{}_{\nu}\,\psi^{\nu}$ in the
second OPE, consistency with the first OPE requires that
\[
(\g^\mu)_A{}^B=D^\mu{}_\nu\,(M^{-1}\,\g^\nu\,M)_A{}^B\ \ .
\]
Alternatively, this identity may be written
as $(M\,\g^\mu)=D^\mu{}_\nu\,(\g^\nu\,M)$. In other words,
$M$ (anti)commutes the $\g^\mu$ when $\mu$ corresponds
to a direction orthogonal (parallel) to the D$p$-brane world-volume.
Hence one concludes
\[
M=\left\lbrace\matrix{
&a\,\g^0\cdots\g^p\,\hphantom{\g_{11}}&\qquad {\rm for}\ p+1\ {\rm odd}\cr
&b\,\g^0\cdots\g^p\,\g_{11}&\qquad {\rm for}\ p+1\ {\rm even}\cr}
\right.
\]
where $a$ and $b$ are undetermined phase factors. 
Note that in the scattering amplitudes the factor of $\g_{11}$
for $p+1$ even simply provides an extra sign in the amplitudes,
and so this form essentially agrees with that in Ref.~\cite{igorb}.
The phase factors $a$ and $b$ may
be further fixed by the following OPE's \cite{danf}
\beqar
S_A(z)\,S_B(w)&\sim&(z-w)^{-\frac{5}{4}}\,C^{-1}_{AB}+\cdots
\\
\tS_A(\bz)\,\tS_B(\bw)&\sim&(\bz-\bw)^{-\frac{5}{4}}\,C^{-1}_{AB}+\cdots\ \ .
\eeqar
In this case consistency with eq.~(\ref{replacedagain}) requires that
\[
M_A{}^C\,M_B{}^D\,C^{-1}_{CD}=C^{-1}_{AB}\ .
\]
Alternatively writing $M\,C^{-1}\,M^T\,C=1$,
we see that $M^{-1}=C^{-1}\,M^T\,C$.
This relation fixes the phase factors to be
$a=\pm i$ and $b=\pm 1$. Hence one may write
\beq
M=\left\lbrace\matrix{
&\frac{\pm i}{(p+1)!}\,(\e^v)_{\mu_0\cdots\mu_p}\,\g^{\mu_0}\cdots\g^{\mu_p}
\,\hphantom{\g_{11}}&\qquad {\rm for}\ p+1\ {\rm odd}\cr
&\frac{\pm 1}{(p+1)!}\,(\e^v)_{\mu_0\cdots\mu_p}\,\g^{\mu_0}\cdots\g^{\mu_p}
\,\g_{11}&\qquad {\rm for}\ p+1\ {\rm even}\ .\cr}
\right.
\labell{finalm}
\eeq
The remaining ambiguity in the sign 
ultimately determines the sign of the R-R charge carried by the
D$p$-brane, as shown in sect.~\ref{back}.
With our choice of conventions, $M$ has the following symmetries
\[
(M)_{AB}=(-1)^{\frac{1}{2}p(p+1)}\,(M)_{BA}
\qquad\quad M^T=-(-1)^{\frac{1}{2}p(p+1)}\,C\,M\,C^{-1}\ \ .
\]

\end{document}